\documentclass[acmsmall,screen]{acmart}
\setcopyright{none}
\renewcommand\footnotetextcopyrightpermission[1]{}
\AtBeginDocument{%
  }
\usepackage{booktabs}
\usepackage{array}
\usepackage{threeparttable}
\usepackage{soul}
\usepackage{tikz}
\usepackage[ruled,longend,linesnumbered]{algorithm2e}
\usepackage{multirow}
\usepackage{subfig}
\usepackage[many]{tcolorbox}
\usepackage{caption}
\usepackage[table, HTML]{xcolor}
\usepackage{arydshln}
\usepackage[normalem]{ulem}
\usepackage{xspace}

\definecolor{mean_color}{HTML}{fde0ef}
\definecolor{median_color}{HTML}{5b3363}
\definecolor{light_green}{HTML}{c8e3e1}
\definecolor{light_red}{HTML}{f5dbe1}

\newcommand {\bdiff} {BDiff\xspace}
\newcommand {\llm} {LLM\xspace}
\newcommand {\Ast} {AST\xspace}
\newcommand {\lcs} {LCS\xspace}

\setcopyright{acmlicensed}
\copyrightyear{2025}
\acmYear{2025}
\acmDOI{XXXXXXX.XXXXXXX}

\acmConference[Conference acronym 'XX]{Make sure to enter the correct
  conference title from your rights confirmation email}{June 03--05,
  2018}{Woodstock, NY}

\acmISBN{978-1-4503-XXXX-X/2018/06}

\begin{document}

\title{\bdiff: Block-aware and Accurate Text-based Code Differencing}

\author{Yao Lu}
\email{luyao08@nudt.edu.cn}
\orcid{0000-0002-3520-5829}
\author{Wanwei Liu}
\email{wwliu@nudt.edu.cn}
\author{Tanghaoran Zhang}
\email{zhangthr@nudt.edu.cn}
\author{Kang Yang}
\email{yangkang@nudt.edu.cn}
\author{Yang Zhang}
\email{yangzhang15@nudt.edu.cn}
\author{Wenyu Xu}
\email{xuwenyu@nudt.edu.cn}
\author{Longfei Sun}
\email{lfsun@nudt.edu.cn}
\author{Xinjun Mao}
\email{xjmao@nudt.edu.cn}
\affiliation{%
  \institution{National University of Defense Technology}
  \city{Chang Sha}
  \state{Hunan}
  \country{China}
}

\author{Shuzheng Gao}
\email{szgao23@cse.cuhk.edu.hk}
\author{Michael R. Lyu}
\email{lyu@cse.cuhk.edu.hk}
\affiliation{%
  \institution{The Chinese University of Hong Kong}
  \city{Hong Kong}
  \country{China}}

\definecolor{myblue}{RGB}{0,112,192}
\renewcommand{\shortauthors}{Lu et al.}

\newcommand*\circled[2][black]{%
  \tikz[baseline=(char.base)]{
    \node[shape=circle, draw=myblue, fill=myblue, inner sep=1pt, text=white, font=\scriptsize\bfseries] (char) {#2};%
  }
}

\begin{abstract}
Code differencing is a fundamental technique in software engineering practice and research. The most widely used differencing technology today remains text-based differencing methods, which operate at the textual line level. While researchers have proposed text-based differencing techniques capable of identifying line changes over the past decade, existing methods exhibit a notable limitation in identifying edit actions (\textit{EA}s) that operate on text blocks spanning multiple lines. Such \emph{EA}s are common in developers' practice, such as moving a code block for conditional branching or duplicating a method definition block for overloading. 
Existing tools represent such block-level operations as discrete sequences of line-level \textit{EA}s, compelling developers to manually correlate them and thereby substantially impeding the efficiency of change comprehension.
To address this issue, we propose \bdiff, a text-based differencing algorithm capable of identifying two types of block-level \textit{EA}s and five types of line-level \textit{EA}s. Building on traditional differencing algorithms, we first construct a candidate set containing all possible line mappings and block mappings. Leveraging the Kuhn-Munkres algorithm, we then compute the optimal mapping set that can minimize the size of the edit script (\emph{ES}) while closely aligning with the original developer’s intent. To validate the effectiveness of \bdiff, we selected five state-of-the-art tools, including large language models ({\llm}s), as baselines and adopted a combined qualitative and quantitative approach to evaluate their performance in terms of \emph{ES} size, result quality, and running time. Experimental results show that \bdiff produces higher-quality differencing results than baseline tools while maintaining competitive runtime performance. Our experiments also show the unreliability of {\llm}s in code differencing tasks regarding result quality and their infeasibility in terms of runtime efficiency. Based on the proposed algorithm, we have implemented a web-based visual differencing tool, which can be integrated into Git. We have open-sourced our tool at \url{https://github.com/bdiff/bdiff}.

\end{abstract}



\received{20 February 2007}
\received[revised]{12 March 2009}
\received[accepted]{5 June 2009}

\maketitle

\section{Introduction}
\label{intro}
Code differencing, a technique that computes the differences between two program versions, is fundamental and essential for both software engineering (SE) practice and research. Developers use code differencing tools to review code changes, debug defects, and track code evolution. Computer-aided software engineering tools rely on code differencing techniques as foundational support for tasks such as code merging. Furthermore, researchers employ these techniques in studies aimed at identifying bug introductions~\cite{da2016framework, hata2012bug, quach2021empirical}, mining change patterns~\cite{martinez2019coming, Tanghaoran2024How}, measuring developer contributions~\cite{gousios2008measuring, Lu2018An}, etc.
\par Existing code-differencing algorithms can be classified into two categories based on the representation of code~\cite{MiryungDiscovering09}: (1) text-based algorithms working on textual lines and (2) tree-based algorithms working on an abstract syntax tree (\Ast). Text-based methods, such as the traditional Myers algorithm~\cite{Myers1986An}, compare raw source code at the character or line level. They typically use techniques like the longest common subsequence (\lcs) to detect basic changes of lines and can be applied to any text file. In contrast, tree-based methods, such as ChangeDistiller \cite{Fluri2007Change} and GumTree~\cite{Falleri2014Fine}, parse code into {\Ast}s and compare these {\Ast}s to detect differences. These approaches capture structural and semantic changes more accurately, offering more fine-grained changes. A major limitation of these methods is that they are language-dependent and require adapting different parsers for various programming languages and language versions. Currently, owing to their universality, efficiency, and robustness, developers generally use text-based differencing tools in practice~\cite{Fluri2021A, Yusuf2019How}, and we focus on this technical approach.
\begin{figure*}[!t]
    \centering
    \includegraphics[width=\linewidth]{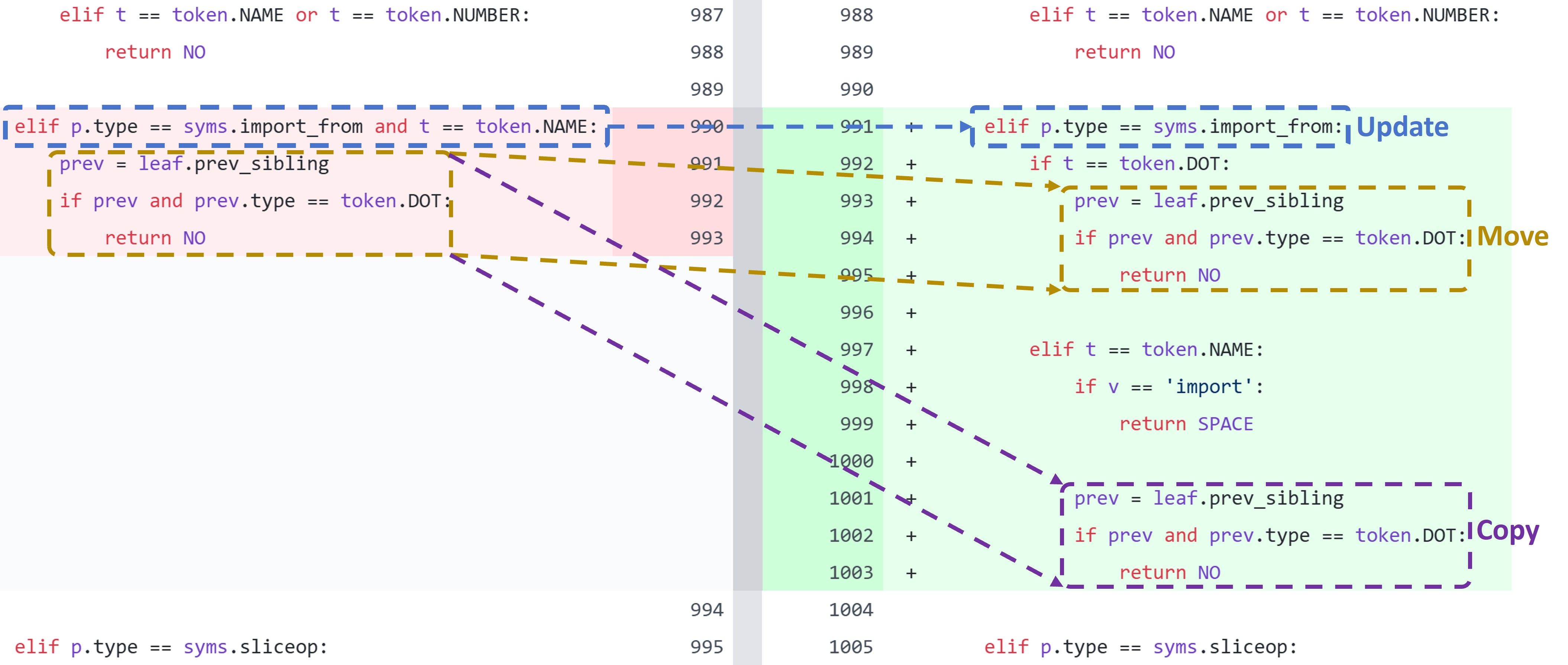}
    \caption{A real case of line updating, block moving, and block copying (psf/black, e1e8909, black.py)}
    \label{block_example}
\end{figure*}
\par Text-based differencing has been a long-standing research topic, and a suite of notable algorithms has been proposed and refined over time. The Hunt-McIlroy algorithm~\cite{Hunt1976An}, proposed in 1976, serves as a foundational method for computing the LCS between two sequences. The Myers algorithm~\cite{Myers1986An}, adopting a greedy strategy, is a highly efficient and influential method for computing the minimum edit script (\emph{ES}) between two sequences, \textit{i.e.}, the minimum number of edit actions (\textit{EA}s) required to transform one sequence into the other. A limitation of these traditional algorithms is that the computed \textit{EA}s are only represented as deletions and insertions. These primitive \textit{EA}s often fail to capture developers' actual changes such as line updates. To address this limitation, Canfora et al.~\cite{GerardoTracking2009, Gerardo2009ldiff} proposed ldiff, a method that takes the output of the Myers algorithm as input and can identify line updates and line moves. Furthermore, other related works~\cite{Godfrey2005Using, Tracking2007Ekwa, Reiss2008Tracking, LHDiffAsaduzzaman2013} have developed algorithms capable of tracking the evolution of source code lines, e.g., line splitting and line merging in LHDiff~\cite{LHDiffAsaduzzaman2013}.
\par While existing works have made progress in identifying line-level changes, they exhibit a key limitation in detecting block-level changes, which poses substantial challenges for developers in effectively understanding actual change intentions. In practice, developers frequently perform edit operations on multi-line code blocks, e.g., moving a code block into an \textit{if}-condition structure for defensive programming, or duplicating a function block with subsequent modifications to implement function overloading~\cite{Ahmed2015An}. Figure~\ref{block_example} presents the Git diff view of a historical commit from the GitHub project ``\textit{psf/black}''. The differencing results consist of 4 consecutive deleted lines and 13 consecutive added lines. Through a detailed analysis, however, we can infer that the developer actually moved a code block (spanning lines 991–993) into an \textit{if}-condition and copied the same block into an \textit{elif}-condition. This example illustrates that a single block-level \textit{EA} corresponds to multiple line-level \textit{EA}s. Identifying such block-level \textit{EA}s can therefore substantially reduce the \textit{ES} size and help developers more efficiently comprehend edit intent. Moreover, in real-world development scenarios, the two blocks involved in block-level \textit{EA}s may be far apart, encompassing both changed code (for block movement) and unchanged code (for block duplication) from the original version, and the identified block-level \textit{EA}s can therefore assist developers in evaluating the impact of code changes. However, identifying both line-level and block-level \emph{EA}s that can minimize the size of the \emph{ES} and are close to the original developer’s intent is challenging. Code files often contain multiple instances of identical or highly similar lines and blocks-this ambiguity can lead to scenarios where a single line or block maps to multiple counterparts in the other version. Computing an optimal mapping set that establishes mappings between lines and blocks across the original and modified code versions remains a non-trivial challenge.
\par To address the challenges, in this paper, we present \bdiff, a novel text-based differencing algorithm which can effectively identify two types of block-level \textit{EA}s and five types of line-level \textit{EA}s. \bdiff is composed of three successive phases. First, building on the results of traditional differencing algorithms, we construct the candidate set containing all possible line mappings and block mappings. Then, we model the mappings as a weighted bipartite graph and iteratively use the Kuhn-Munkres algorithm~\cite{Munkres1957AlgorithmsFT} to compute the optimal mapping set, which can minimize the edit-script size and are close to the original developer intent. Finally, based on the optimal mapping set, we deduce the \emph{ES}. To validate \bdiff’s effectiveness, we constructed an evaluation dataset consisting of 2,997 real-world code-change cases, covering three popular languages: Python, Java, and XML. We selected five baseline tools representing three distinct technical paradigms (text-based, AST-based, and large language model (LLM)-based approaches) and quantitatively compared {\bdiff}’s performance against these baselines in \textit{ES} size and running time. To investigate the quality of {\bdiff}'s results, we further conducted two experiments: a qualitative manual evaluation experiment involving 10 raters and a mutation-based evaluation experiment. The experimental results show that: (1) \bdiff outperforms all baseline tools in differencing result quality; (2) \bdiff outperforms text-based baseline tools in \textit{ES} size, reducing the average \textit{ES} size by at least 28\%; (3) \bdiff outperforms all baseline differencing tools in running time (0.085s on average), with the only exception being Git diff (Myers); and (4) the accuracy of \bdiff in identifying ground-truth \textit{EA}s and \textit{ES} are 95.3\% and 82.4\%, respectively. Based on our algorithm, we have implemented an open-source and web-based diff visualization tool. To summarize, our work makes the following contributions:
\begin{itemize}
    \item To the best of our knowledge, we are the first to explore the identification of block-level \textit{EA}s which are common in developers' edit operations.
    \item We propose \bdiff, a systematic approach to identify both line-level \textit{EA}s and block-level \textit{EA}s, which consists of three phases: establishing cross-version mappings, computing an optimal mapping set, and deducing the \emph{ES}.
    \item We conduct an extensive experiment to evaluate the performance of \bdiff and baseline tools. Experimental results show that BDiff produces higher-quality differencing results than baseline tools while maintaining competitive runtime performance. 
    \item We have implemented an open-source visual differencing tool, which can be integrated into Git.
\end{itemize}
\par Our dataset, scripts, and experimental results are available in the GitHub repository\footnote{https://github.com/\bdiff/\bdiff-Evaluation-Experiment}.
\section{Related Work}
\label{rw}
\subsection{Text-based Code Differencing}
\label{rw1}
Traditional text-based differencing techniques typically compare two files at the line granularity, with outputs consisting of line insertions and deletions. Hunt and McIlroy introduced a foundational method for computing text file differences—the seminal \textit{Hunt-McIlroy} algorithm~\cite{Hunt1976An}. This algorithm employs dynamic programming to efficiently identify the LCS between files. It was employed as the early differencing algorithm in \texttt{Unix diff}. Another prominent line-differencing algorithm is Myers~\cite{Myers1986An}, which uses dynamic programming to find a LCS with a time complexity of O(N+$D^2$). It superseded the \textit{Hunt-McIlroy} algorithm as the differencing algorithm for Unix \texttt{diff} and has been widely adopted in version control systems such as Git. Besides the default Myers, Git offers three alternative diff algorithms: \textit{Minimal}, \textit{Patience}, and \textit{Histogram}. Notably, \textit{Minimal} and \textit{Histogram} are improved variants of Myers and \textit{Patience}, respectively. Through a systematic mapping study, Nugroho et al.~\cite{Yusuf2019How} observed that Git's three alternative diff algorithms have seen limited adoption in existing works. They further investigated the impact of Myers and \textit{Histogram} on three major applications: code churn metrics, the \textit{SZZ} algorithm~\cite{Jacek2005When}, and patch extraction. Their findings revealed that \textit{Histogram} exhibits superior performance in describing code changes.
\par Over the past decade, numerous studies have explored the use of differencing algorithms for tracking code locations. Godfrey and Zou~\cite{Godfrey2005Using} introduced a method to detect merging and splitting of source code entities. Reiss~\cite{Reiss2008Tracking} conducted a comprehensive evaluation of methods for maintaining source locations, finding that \textit{W\_BESTI\_LINE}, which uses Levenshtein distance for string comparison, incorporates context lines, and ignores indentation, matches the effectiveness of other methods while offering speed and low storage requirements. Canfora et al.~\cite{GerardoTracking2009, Gerardo2009ldiff} proposed ldiff, a differencing tool that can track line updates and movements. Based on the results of Unix \texttt{diff}, it compares similarity across all possible hunk pairs between original and new versions. For the most similar hunks, it further analyzes individual lines to detect changes. Asaduzzaman et al.~\cite{LHDiffAsaduzzaman2013} proposed \textit{LHDiff}, which employs locality-sensitive hashing and normalized Levenshtein edit distance to compute content similarity by considering both content and context.
\par Despite these advancements, current text-based differencing techniques still suffer from a significant limitation: they are incapable of identifying block-level \textit{EA}s, such as block moves and copies, which are routine in developers' coding practices. In this paper, we propose a text-based differencing algorithm capable of identifying block-level \textit{EA}s, as well as five types of line-level \textit{EA}s.
\subsection{Tree-based Code Differencing}
Since source code can be represented as {\Ast}s, tree differencing techniques can be used to compute \Ast differences and derive fine-grained code changes. Hashimoto and Mori \cite{Hashimoto2008Diff} proposed Diff/TS, a tool that visualizes \textit{EA}s through detailed structural analysis of source code trees. Spacco et al. \cite{Spacco2009Lightweight} introduced SDiff, a hybrid differencing technique that combines line-based and {\Ast}-based approaches. Building on an existing tree differencing algorithm~\cite{Chawathe1996Change}, Fluri et al. \cite{Fluri2007Change} developed ChangeDistiller to extract fine-grained source code changes.
\par Among various AST-based code differencing algorithms, GumTree~\cite{Falleri2014Fine} is an influential one. Its node-mapping process consists of two phases: 1) a greedy top-down phase to find isomorphic subtrees, and 2) a bottom-up phase to match tree nodes. Evaluation results demonstrate that GumTree's differencing outputs are often more comprehensible than those of Unix diff. To address the limitation of GumTree in handling large code files, Falleri et al.~\cite{Falleri2024Fine} proposed gumtree-simple, an improved heuristic of GumTree that is not only significantly faster but also produces smaller and more understandable \textit{ES}s. In recent years, many studies on AST-based differencing have been carried out based on GumTree. Based on GumTree, Matsumoto et al.~\cite{atsumoto2019Beyond} leveraged both AST and line differences to generate more interpretable \textit{ES}s. Building on GumTree, Frick et al.~\cite{Frick2018Generating} introduced IJM, a differencing approach that can generate more accurate and compact \textit{ES}s capturing developers' intent. Dotzler and Philippsen~\cite{Dotzler2016Move} proposed MTDIFF that refines the heuristics of GumTree to find more move actions. Building on GumTree, Dilavrec et al.~\cite{Quentin2022HyperAST,DilavrecHyperDiff2023} proposed HyperDiff, an approach that addresses the scalability issue of computing diffs over large code histories.
Huang et al.~\cite{Huang2018ClDiff} proposed ClDiff that can generate concise, linked, and understandable code differences with a granularity between code differencing and code change summarization methods.
Fan et al.~\cite{Fan2021An} proposed a hierarchical approach to automatically compare the similarity of mapped statements and tokens across different AST mapping algorithms. Their experimental results reveal that state-of-the-art AST mapping algorithms still produce a considerable number of inaccurate mappings.
\par As evident from the preceding review of relevant studies, recent research on code differencing has primarily focused on AST-based approaches. Compared with text-based differencing algorithms, AST-based methods can extract fine-grained, syntax-aware changes, and scholars often leverage such change information for research purposes~\cite{Fluri2021A}. In the code review scenario, however, developers typically need to first grasp the information of textual change at the surface-level. Furthermore, tree-based differencing approaches rely on language-specific parsers~\cite{Falleri2014Fine}; yet modern software projects often involve multiple programming languages, and these languages undergo continuous evolution. This introduces practical barriers in real-world software development contexts and may even lead to analysis failures. For these reasons, traditional text-based differencing tools continue to be widely used in real-world development practices~\cite{German2019cregit}.

\section{The \bdiff Algorithm}
\subsection{Approach Overview}
\label{AO}
\begin{figure}[!t]
    \centering
\includegraphics[width=\linewidth]{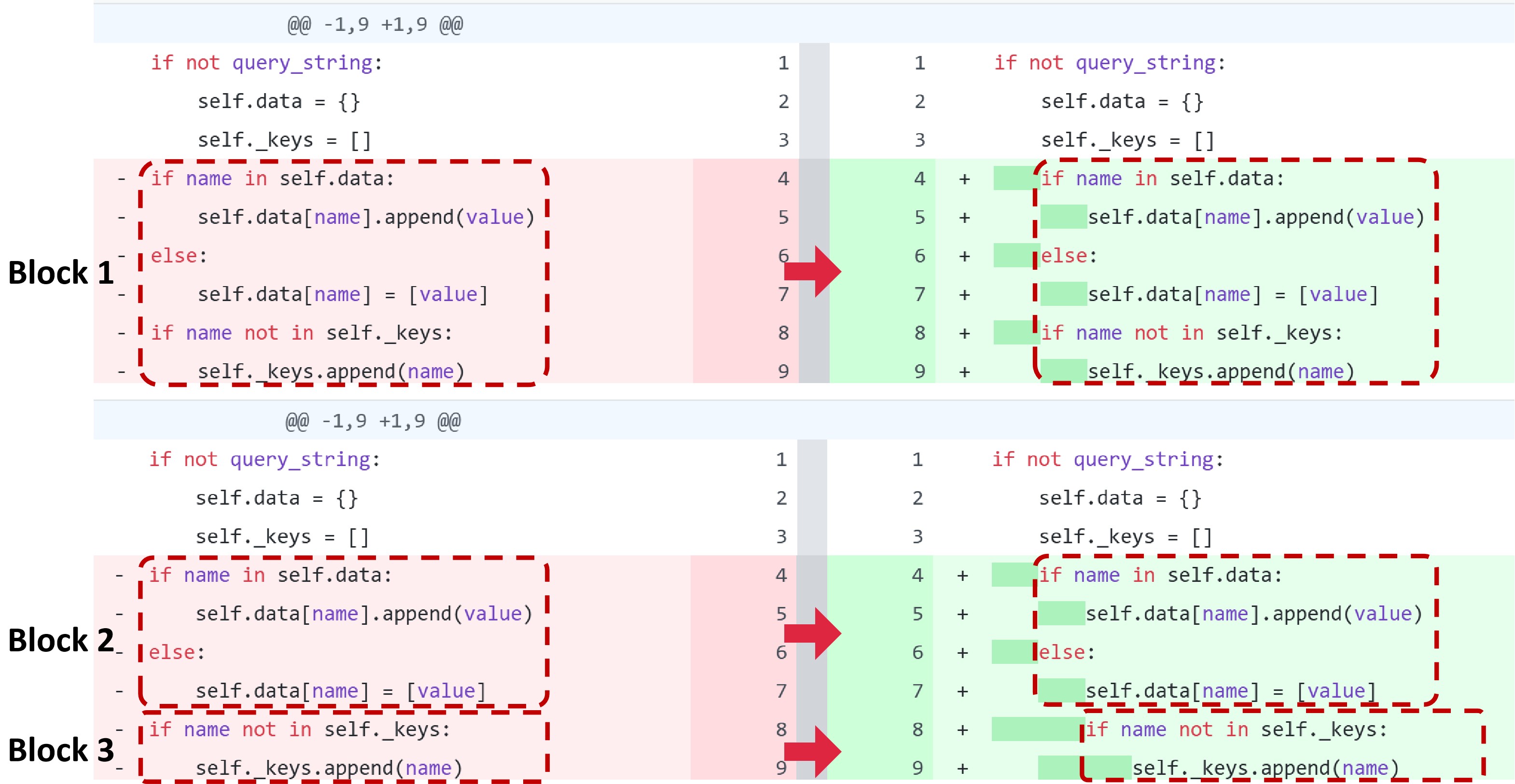}
    \caption{Examples of block-level EAs}
    \label{block-example}
\end{figure}
Before developing an effective code differencing algorithm, we first need to figure out what constitutes a high-quality code differencing result. Regarding this question, there is a well-established consensus in prior research~\cite{Kim2005An, Yusuf2019How}: (1) from an internal perspective, the result should have a minimal \textit{ES} to reduce the time developers spend reading changed code; and (2) from an external perspective, it should systematically present changes to help developers understand the actual changes made and their impacts on the code. This leads us to a second question that we need to clarify: What types of \textit{EA}s do developers perform during code editing? From the perspective of \textit{EA} granularity, we classify these \textit{EA}s into two categories: line-level \textit{EAs} and block-level \textit{EAs}. Line-level \textit{EA}s operate on individual lines of code (LOC) and include the following types:
\begin{itemize}
    \item \textbf{Line Deleting (LD)}: removing an entire line.
    \item \textbf{Line Adding (LA)}: inserting a new line.
    \item \textbf{Line Updating (LU)}: modifying a portion of a single line, including three specific scenarios: deleting a segment of the line, inserting a segment into the line, or replacing one segment with another.
    \item \textbf{Line Splitting (LS)}: dividing a single line of code into multiple lines.
    \item \textbf{Line Merging (LM)}: combining multiple LOCs into a single line.
\end{itemize}   
\par \dashuline{\textit{Block-level EAs}} refer to the \textit{EA}s that operate on a set of consecutive lines as a whole, with the relative indentation between lines within the block remaining unchanged. They include two types:
\begin{itemize}
    \item \textbf{Block Moving (BM)}: relocating a contiguous code block to another position in the same file with optional indentation adjustments or only adjusting its indentation level, e.g., moving a code block into a if-condition block for defensive programming.
    \item \textbf{Block Copying (BC)}: duplicating a contiguous code block to a new location in the file with optional indentation adjustments. e.g., duplicating and modifying a method definition block for overloading. Prior research~\cite{Ahmed2015An} has shown that 64\% of copy-and-paste operations in software development occur within a single source file.
\end{itemize}
As illustrated in Figure~\ref{block-example}, the lines 4–9 in \textit{Block 1} are all indented right by one level (equivalent to four spaces), which constitutes a \textit{BM}. In contrast, the lines in \textit{Block 2} and \textit{Block 3} are also indented right, but with different indentation levels, which results in their classification as two separate \textit{BM}s. 
The minimum number of non-blank lines required to form a code block is configurable, with a default value of two. Notably, unlike prior research on source location tracking, we do not identify one-line moves as \textit{EA}s, because they are prone to false positives. Additionally, block-level \textit{EA}s are frequently accompanied by \textit{LU}s. For example, when duplicating a method block to implement overloading, developers typically modify parameters or return types within the copied block. To address this common scenario, \bdiff supports the identification of composite block-level \textit{EA}s, i.e., block-level \textit{EA}s combined with intra-block \textit{LU}s.
\begin{figure}[!t]
    \centering
    \includegraphics[width=\linewidth]{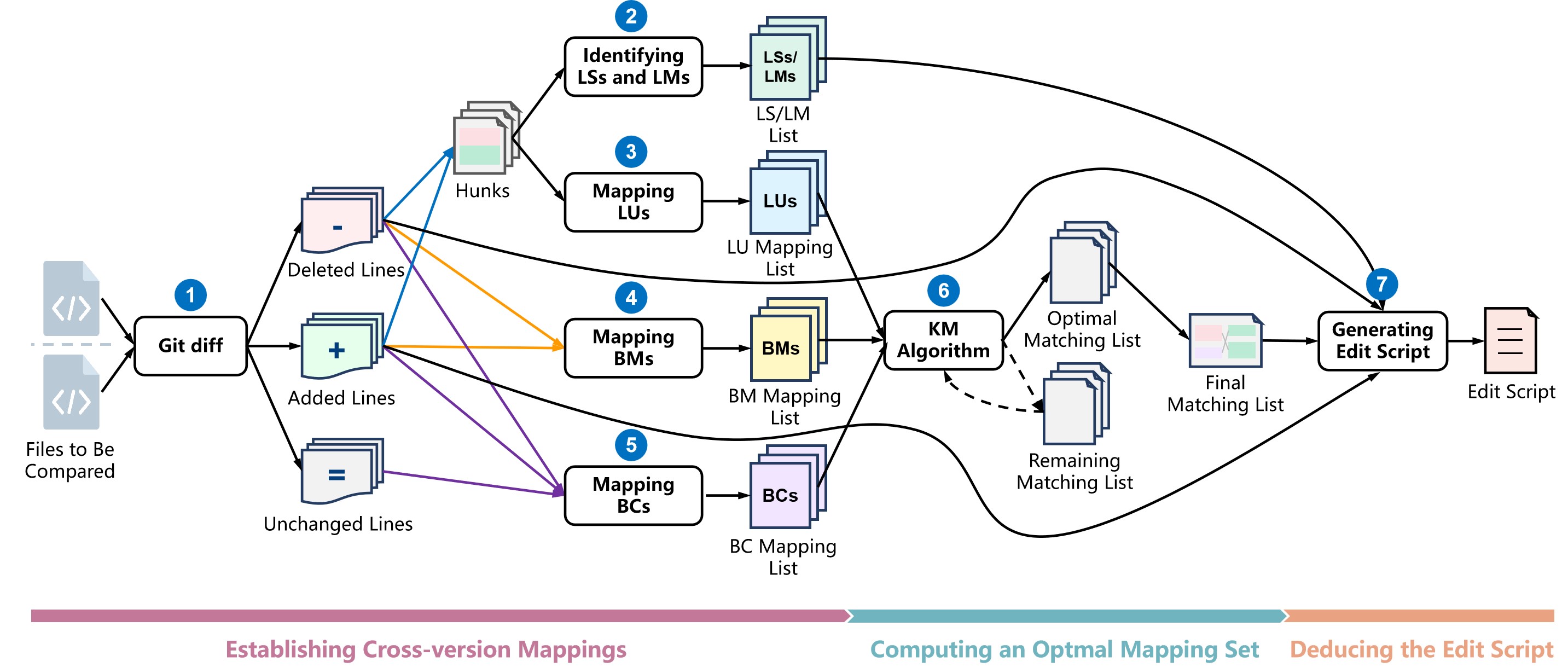}
    \caption{The \bdiff algorithm framework}
    \label{me_fr}
\end{figure}
\par Figure~\ref{me_fr} presents an overview of \bdiff. The inputs to \bdiff are the two versions of a source code file to be compared. From now on, we refer to the original and modified versions as the left and right versions, respectively. Overall, \bdiff is composed of three phases. First, we establish line-level and block-level mappings between the left and right versions. We apply the Git diff command to extract the deleted, added, and unchanged lines~\circled{1}, as prior studies \cite{Reiss2008Tracking, LHDiffAsaduzzaman2013} have demonstrated its high accuracy and efficiency for this task. The deleted and added lines are in the left and right versions, respectively. Since Git diff can be configured to use four different diff algorithms (as introduced in Section~\ref{rw1}), we evaluate the performance of \bdiff on our dataset using each of these algorithms to identify the optimal one. Within each hunk (i.e., a maximal sequence of consecutive deleted and/or added lines describing closely related changes) generated by Git diff, we identify \textit{LS}s and \textit{LM}s and remove the corresponding lines from the original deleted/added line lists and their associated hunks~\circled{2}. For the resulting hunks, we then compare the remaining deleted and added lines to produce a \textit{LU} mapping list~\circled{3}. By comparing all deleted and added lines across the entire file, we obtain a \textit{BM} mapping list~\circled{4}. Since copy actions can originate from either deleted lines (i.e., copying followed by deletion) or unchanged lines, we compare all lines in the left version with the added lines in the right version to obtain a \textit{BC} mapping list~\circled{5}. The mappings within and across the three lists (\textit{LU}, \textit{BM}, and \textit{BC}) may overlap. In the second phase, we therefore synthesize these lists and iteratively apply the Kuhn-Munkres (KM) algorithm to find the optimal final matching set, which includes \textit{LA}s, \textit{BM}s, and \textit{BC}s~\circled{6}. Finally, using the identified \textit{LS}s and \textit{LM}s, the final matching set, and the remaining \textit{LA}s and \textit{LD}s as inputs, we deduce the final \textit{ES}~\circled{7}.
\subsection{LSs and LMs Identification}
\textit{LS}s and \textit{LM}s occur within individual hunks. Since these two \textit{EA} types are reciprocal and symmetric, we only detail the detection process for \textit{LS}s: a single line in the left version is exactly equal to the concatenation of several consecutive lines in the right version.
Unlike \textit{LHDiff}~\cite{LHDiffAsaduzzaman2013}, which relies on text similarity and a threshold to identify "possible" \textit{LS}s that may include line modifications, our approach identifies \textit{LS}s and \textit{LM}s that are "exact" splits or merges, with no additional changes, to ensure high identification accuracy.
To achieve this, we repeatedly check whether the current left line starts with the content of the current right line. If it does, and the two lines are not identical, we update the left line by removing the matched prefix and continue checking against the next right line in the hunk. If, at any point, the processed left line exactly matches the current right line, we have found a \textit{LS}, i.e., the original left line is mapped to all the right lines involved in this process.
We set a maximum number of detection attempts for \textit{LS}s and \textit{LM}s, with a default value of 8~\cite{LHDiffAsaduzzaman2013}, which is configurable by the user. Because \emph{LS}s and \emph{LM}s exhibit a strong correlation between the left and right lines, with a high probability of representing the actual \textit{EA}s, we directly identify them as the final \emph{EA}s. Therefore, once a \textit{LS} or \emph{LM} is found, the corresponding lines are removed from the deleted/added line lists and their associated hunks, so they are not considered in subsequent steps. If a right line is blank, we simply move to the next right line without incrementing the detection attempt counter.

\subsection{Max Intersection Removal-based LU mapping}
\label{LU_sec}
Like \textit{LS}s and \textit{LM}s, \textit{LU}s occur within hunks. For each hunk remaining after the \textit{LS}/\textit{LM} identification phase, we compare each deleted line with each added line to identify all potential \textit{LU} mappings. We adopt the \textit{W\_BESTI\_LINE} method~\cite{Reiss2008Tracking} to determine whether a deleted line and an added line form a \textit{LU} mapping (see Algorithm~\ref{LU_mapping_algo}). This technique combines content and context similarity to track source lines, with prior studies~\cite{Reiss2008Tracking, LHDiffAsaduzzaman2013} validating its effectiveness and efficiency.
\begin{algorithm}[!t]
\footnotesize
    \caption{The LU mapping algorithm}  
    \label{LU_mapping_algo}
    \KwData{A list of lines in the left version $\mathcal{L}$, a list of lines in the right version $\mathcal{R}$, a list of successive deleted lines $\mathcal{D}$ and a list of successive added lines $\mathcal{A}$ within a hunk, context length $ctxLen$, line similarity weight $lineWgt$, combined similarity threshold $simThres$, and an empty LU-mapping dictionary $\mathcal{M}$}
    \KwResult{The LU-mapping dictionary $\mathcal{M}$}
         \If{$\mathcal{D} \neq \emptyset \  \&\& \  \mathcal{A} \neq \emptyset $}{
\ForEach{ $r \in \mathcal{A}$}{
    \ForEach{$l \in \mathcal{D}$}{
         $s \leftarrow W\_BESTI\_LINE(l, r, \mathcal{L}, \mathcal{R}, ctxLen, lineWgt)$\;
   		\If{$ s \geq  simThres$}{$\mathcal{M}[(l, r, s)] \leftarrow \emptyset;$}
    }
}
\ForEach{ $(l_1, r_1, s_1) \in \mathcal{M}$}{
	\ForEach{ $(l_2, r_2, s_2) \in \mathcal{M}$}{
		\If{$(l_1 - l_2) * (r_1 - r_2) < 0$}{$add(\mathcal{M}[(l_1, r_1, s_1)], (l_2, r_2, s_2))$\;}
	}
}
\tcc{Sort $\mathcal{M}$ in ascending order, where the primary key is the number of intersections, and the secondary key is the combined similarity score.}
$\mathcal{M}.sort(sortKey=(size(\mathcal{M}.value), \mathcal{M}.key[2]))$\;
\While{$\mathcal{M} \neq \emptyset $}{
		\If{$\mathcal{M}[-1].value \neq \emptyset $}
		{
			$lastLU \leftarrow  pop(\mathcal{M}[-1])$\;
			\ForEach{$m \in \mathcal{M}$}{
				\If{$lastLU.key \in \mathcal{M}[m]$}
				{
					$\mathcal{M}[m].remove(lastLU.key)$\;
				}
			}
		}
		\If{$\mathcal{M}\ does\ not\ contain\ intersections$}
              {$break$\;}
        $\mathcal{M}.sort(sortKey=(size(\mathcal{M}.value), \mathcal{M}.key[2]))$\;
	}
}
\end{algorithm}
In our implementation, content similarity is measured using Levenshtein ratio, while context similarity is computed as the proportion of matched context lines relative to all context lines in a window. The context window includes 4 lines before and after the target line. The combined similarity score for each line pair is calculated as 0.6 × content similarity + 0.4 × context similarity~\cite{Reiss2008Tracking, LHDiffAsaduzzaman2013}; line pairs with a combined score of at least 0.5 are added to the \textit{LU} mapping list. All parameters mentioned, including context window size, content similarity weight, and combined score threshold, are configurable. Through this approach, candidate \textit{LU} mappings may intersect within a single hunk, which contains line movements and thus violates our constraints. To resolve this issue while maximizing the number of valid  \textit{LU}s extracted, we calculate the intersection count for each mapping and sort mappings in descending order of their intersection counts. We then iteratively remove the mapping with the highest intersection count, update the intersection counts of remaining mappings, and repeat until no intersecting mappings remain. The resulting list constitutes the final \textit{LU} mapping list. In the mapping list, the mappings may contain conflicts where multiple mappings share the same line from either the left or the right version. These conflicts will be systematically resolved in the subsequent \textit{KM} step~\circled{6}.

\subsection{Mapping BMs and BCs}
\par We identify the \textit{BM} mapping list by comparing consecutive added lines with consecutive deleted lines. Since developers may first copy a block and then delete the original lines, we derive the \textit{BC} mapping list by comparing consecutive added lines with all consecutive lines in the left version.
We determine whether a ``left'' line and a ``right'' line matches when they are identical or their Levenshtein ratio exceeds 0.6~\cite{Reiss2008Tracking, LHDiffAsaduzzaman2013}. We record line pairs that are not identical, i.e., the \textit{LU}s associated with \textit{BM}s or \textit{BC}s.
We allow that a block starts with a blank line and contain blank lines, which aligns with actual developer practice and can reduce the \emph{ES}. Because developers typically perform \textit{BM}s and \textit{BC}s on meaningful code, we exclude lines containing only stop words (e.g., `\{`) from the block-size calculation by default. Each time a \textit{BC} or \textit{BM} is found, we compute a weight for it; the details are described in Section~\ref{multiple_mappings_sec}.

\subsection{KM-based Optimal Mapping Computing}
\label{multiple_mappings_sec}
\begin{figure}[!t]
    \centering
    \includegraphics[width=\linewidth]{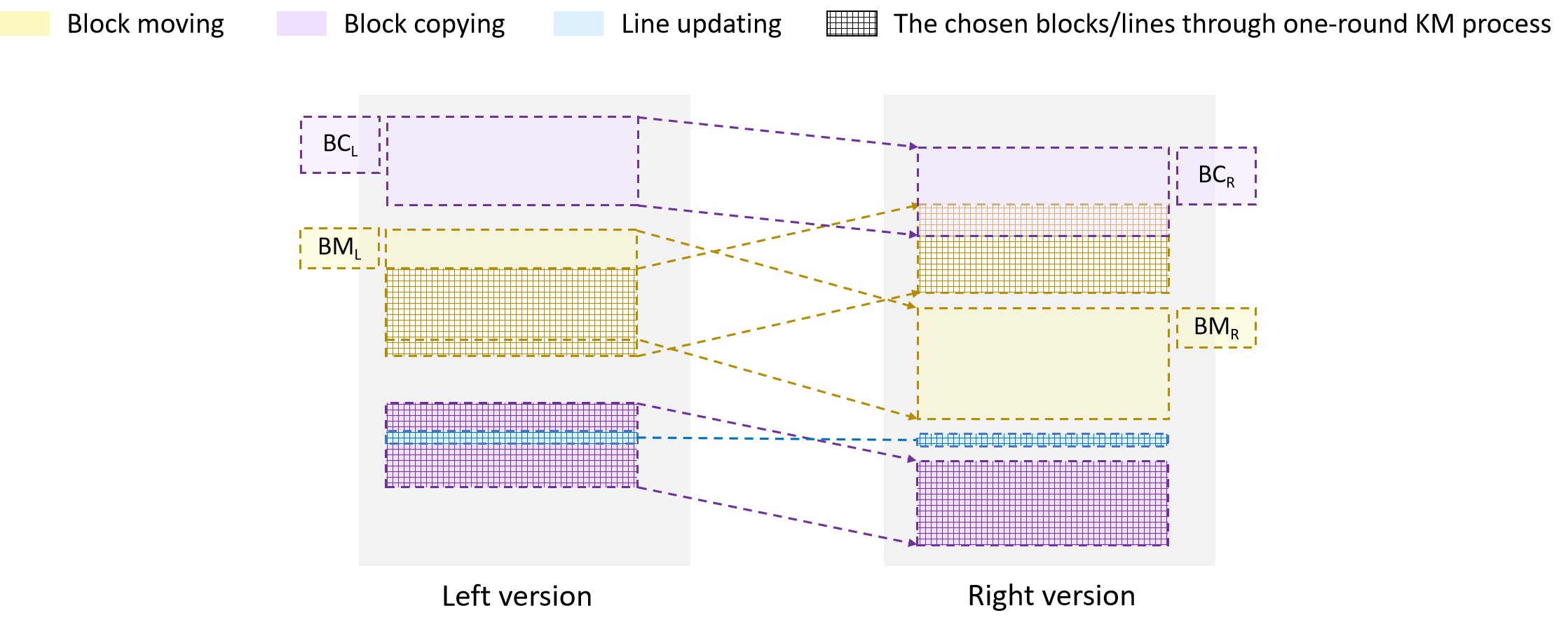}
    \caption{The multiple mapping problem}
    \label{multi_mapping}
\end{figure}

Through the previous two procedures, we obtain three lists of candidate mappings: \textit{LU}s, \textit{BM}s, and \textit{BC}s. These mappings may overlap: both internally within a list, and externally, among the three lists (as illustrated in Figure~\ref{multi_mapping}). However, to form a valid \textit{ES}, each deleted line in the left version except for \textit{BC} and each added line in the right version must be assigned to exactly one source and one target, respectively. To resolve these conflicts, we merge the three mapping lists into a single list, and iteratively apply the KM algorithm~\cite{Kuhn1955TheHM, Munkres1957AlgorithmsFT} to find the optimal matching set. The KM algorithm is an efficient combinatorial optimization method for finding the maximum or minimum weight matching in a bipartite graph, ensuring an optimal one-to-one correspondence between two disjoint vertex sets based on edge weights. Aligning with the characteristics of good differencing results outlined in Section~\ref{AO}, our objective is to minimize the \textit{ES} size while preserving the original edit intent as much as possible.
\par We model the block and line mappings as a weighted bipartite graph $G = (V, E)$, where $V$ is the set of vertices (i.e., the mapped blocks or lines) and $E$ is the set of edges (i.e., the mapping relations). The vertex set $V$ is partitioned into two disjoint subsets, $V_1$ and $V_2$, where $V_1$ contains the mapped blocks or lines from the left version and $V_2$ contains those from the right version. Each edge $e \in E$ is defined according to Equation~(\ref{edge}).
\begin{equation}
\label{edge}
e = (u, v, w)
\end{equation}
where $u \in V_1$, $v \in V_2$, and $w$ is the weight assigned to the vertex pair $(u, v)$. Using this weighted bipartite graph model, our problem reduces to finding a minimum weight matching, which can minimize the \textit{ES} size while maximally reflect the original edit intent. To this end, we calculate the weight $w^{(b)}_{e}$ for \textit{BM} or \textit{BC} using Equation~(\ref{edge_weight}).
\begin{equation}
\footnotesize
\label{edge_weight}
w^{(b)}_{e} = \frac{EditTimes(e)}{Len(u)} + \frac{1-CtxSim(u, v)}{10} + \frac{Dist(u,v)}{100}
\end{equation}
where:
\begin{itemize}
    \item $EditTimes(e)$ computes the number of edits required for mapping $e$. When other factors are equal, mappings with fewer edits are prioritized during the conflict resolution in the KM process, because they are associated with shorter \textit{ES} and are more likely the actual \textit{EA}s. We initialize the values for \textit{BM} and \textit{BC} mappings to 2 and 3, respectively, as the former requires fewer edits than the latter. Since \textit{BM}s or \textit{BC}s may be accompanied by \textit{LU}s and indentation changes, the value is incremented by 1 if they involve indentation changes or are associated with a \textit{LU}.
    \item $Len(u)$ calculates the LOC of $u$ (or $v$) in the mapped block. When resolving overlapping vertices with otherwise equivalent weights, priority is given to the vertex with greater block length.
    \item $CtxSim(u, v)$ measures the context similarity between $u$ and $v$. For overlapping vertices, higher context similarity indicates a stronger alignment with the original edit intent. We measure the context similarity by calculating the Levenshtein ratio of the 4 lines above and below both $u$ and $v$.
    \item $Dist(u, v)$ quantifies the relative line distance between $u$ and $v$. Based on the principle of locality, where programs tend to reuse data and instructions near recently accessed ones, closer mappings are more likely to reflect the original edit intent. This metric is derived from Git diff results, calculated as the sum of unchanged lines plus the maximum of added and deleted lines between the two start lines of the $u$ and $v$.
\end{itemize}
\par Overall, we designed the weight calculation method in this way to reflect the priority of the three components, where the order is: block length > context similarity > relative line distance. This is because, in most cases, the orders of magnitude of the first, second, and third components in Equation ~(\ref{edge_weight}) are 0.1, 0.01, and 0.001, respectively. Similarly, we calculate the weight $w^{(l)}_{e}$ for \textit{LU} using Equation~(\ref{lu_edge_weight}). Since \textit{LU}s inherently occur within a single hunk, we do not include the distance factor when calculating their weight.
\begin{algorithm}[!t]
\footnotesize
    \caption{Handling multiple mappings based on the KM algorithm}
    \label{multiple_mapping_algo}
    \KwData{A \textit{LU} mapping list $\mathcal{LU}$, A \textit{BM} mapping list $\mathcal{BM}$, A \textit{BC} mapping list $\mathcal{BC}$, a list of lines in the left version $\mathcal{L}$, a list of lines in the right version $\mathcal{R}$, minimal BM block length $minBM$, and minimal \textit{BC} block length $minBC$}
    \KwResult{The final optimal matching list $\mathcal{M}$}
	$M_a \leftarrow \mathcal{LU} \cup \mathcal{BM} \cup \mathcal{BC}$\;
	 \If{$M_a \neq \emptyset$}{
	$M_k, M_r \leftarrow km(M_a, \mathcal{L}, \mathcal{R}, minBM, minBC) $\;
	$\mathcal{M} \leftarrow \mathcal{M} \cup M_k$\;
	\While{$M_r \neq \emptyset$}{
		$M_k, M_r \leftarrow km(M_r, \mathcal{L}, \mathcal{R}, minBM, minBC) $\;
		$\mathcal{M} \leftarrow \mathcal{M} \cup M_k$\;		
	}
}
\end{algorithm}
\begin{equation}
\footnotesize
\label{lu_edge_weight}
w^{(l)}_{e} = \frac{EditTimes(e)}{Len(u)} + \frac{1-Sim(u, v)}{10}
\end{equation}
where:
\begin{itemize}
    \item $EditTimes(e)$ is set to 1, as \textit{LU} mappings have an initial value of 1 and involve no additional edits.
    \item $Len(u)$ equals 1, since the \textit{LU} consists of a single line of code.
    \item $Sim(u,v)$ corresponds to the \textit{W\_BESTI\_LINE} value~\cite{Reiss2008Tracking} associated with mapping $e$.
\end{itemize}
\par When constructing the bipartite graph, we treat left overlapping vertices that share deleted lines or right overlapping vertices that share added lines as a single vertex. As illustrated in Figure~\ref{multi_mapping}, the two left \textit{BM} vertices overlap and thus form one left vertex; similarly, a right \textit{BM} vertex and a right \textit{BC} vertex that overlap form one vertex. Because \textit{BC} can be superimposed with other \textit{EA}s, we treat the left vertices of \textit{BC}s as additional vertices. For instance, in Figure~\ref{multi_mapping} the left vertex of \textit{LU} and the left vertex of the \textit{BC} at the bottom that overlap are two distinct vertices. However, this approach can lead to a fragmentation issue: after one round of the KM process, several non-overlapping ``pieces'' of the unselected blocks or lines may remain. These pieces still need to be considered for inclusion in the \textit{ES}, such as the mappings $BC_L \rightarrow BC_R$ and $BM_L \rightarrow BM_R$ in Figure~\ref{multi_mapping}. Note that these remaining pieces may still overlap. To address this, we iteratively apply the KM algorithm to the remaining pieces to find an optimal matching in each round, adding the results to the final matching set until no pieces remain (Algorithm~\ref{multiple_mapping_algo}).

\section{The \bdiff Tool}
\begin{figure}[!t]
    \centering
\includegraphics[width=\linewidth]{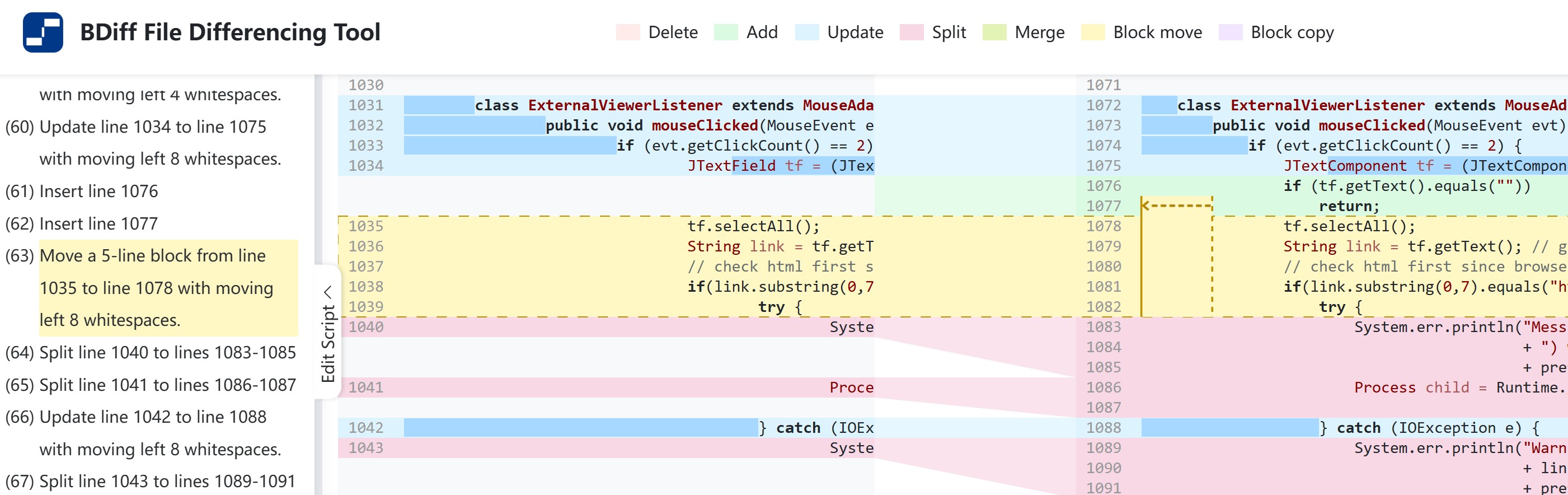}
    \caption{A screenshot of BDiff webpage}
    \label{bdiffscreen}
\end{figure}
Based on the algorithm described in the previous section, we have implemented a web-based, open-source visual differencing tool~\footnote{https://github.com/bdiff/bdiff}. Our tool allows developers to specify the \textit{EA}s to be identified and to configure algorithm-related parameters, e.g., tab size, the synthetic line-similarity threshold, the minimum block lengths for \textit{BM} and \textit{BC}, etc. To facilitate developers in reviewing differences, our tool supports bidirectional navigation between differences and the \textit{EA}s in the \textit{ES} (see Figure~\ref{bdiffscreen}). We offer five ways to use the \bdiff tool: (1) through the website\footnote{http://www.bdiff.net/}, where developers can select two files to compare via a browser and view the results; 
(2) through Git, by configuring \bdiff as the \textit{difftool}\footnote{https://github.com/BDiff/BDiff-Git} and viewing results via Git commands; (3) by exporting the differencing result page using our program\footnote{https://github.com/bdiff/bdiff-Visualization-Exporter}; and (4) by directly obtaining the \textit{ES} data through the provided API\footnote{https://s.apifox.cn/aa189f38-f9da-4ff2-9e08-bc65663b7708/api-198645561}.

\section{Evaluation}
\subsection{Evaluation Setup}
We now present the empirical evaluation of \bdiff. Our goal is to assess its performance against existing state-of-the-art code differencing tools. We first focus on \textit{ES} size, a key quality metric; shorter \textit{ES} sizes generally reduce the time developers need to understand the changes~\cite{Falleri2014Fine}, aligning with our first criterion for good differencing results (Section~\ref{AO}). This leads to our first research question:\\
\textit{\textbf{RQ1}}: Does \bdiff produce smaller \textit{ES} sizes than existing code differencing tools?
\par Another evaluation criterion for a differencing tool is whether it helps developers efficiently understand the actual changes made and their impacts on the code, i.e., the perceived quality of the differences. Thus, our second research question is:\\
\textit{\textbf{RQ2}}: Is the perceived quality of \bdiff's output higher than that of existing code differencing tools?
\par Through {\textit{RQ2}, we assess the perceived quality of \bdiff's results relative to other tools. Additionally, we want to evaluate the correctness of \bdiff's results, i.e., the degree to which its output matches the original edits:\\
\textit{\textbf{RQ3}}: What is the correctness of the \textit{ES}s computed by \bdiff?
\par The previous research questions focus on functional aspects. Our fourth question evaluates runtime performance:\\
\textit{\textbf{RQ4}}: How does \bdiff's runtime performance compare with existing code differencing tools?
\par To answer these questions, we used a mixed-method approach that combined qualitative and quantitative methods. For \textit{RQ1} and \textit{RQ4}, we conducted quantitative analyses of relevant metrics. For \textit{RQ2}, we performed a manual evaluation with 10 raters to collect qualitative feedback. For \textit{RQ3}, we implemented an algorithm that generates random changes to code files, producing ground-truth \textit{ES}s, and then we compared the \textit{ES}s computed by \bdiff against the ground-truth \textit{ES}s.
\par We selected five baseline tools:
(1) Git diff (2.41.0.windows.1, default Myers algorithm)~\cite{gitdiff}, the most widely-used text-based differencing tool, 
(2) ldiff (1.0.8)~\cite{ldiff}, a text-based differencing tool that can identify line updates and moves, 
(3) GumTree (v4.0.0-beta4, in which \textit{Simple}~\cite{Falleri2024Fine} is the default matcher)~\cite{GumTree}, the most advanced state-of-the-art tree-based differencing tool,
(4) GPT-5-mini, a recent closed-source LLM~\cite{gpt5-mini}, and
(5) and Qwen3-32B, an advanced open-source LLM~\cite{qwen3-32B}. We prompt the two LLMs to generate \textit{ES}s in the same \textit{EA} types and format with \bdiff. LHDiff~\cite{LHDiffAsaduzzaman2013} was excluded as it produces line mappings for the line-tracking scenario rather than an \textit{ES}. In the \bdiff algorithm, we use Git diff (2.41.0.windows.1) to obtain the deleted and added lines~\circled{1}.
\subsubsection{Dataset}
We evaluated \bdiff's performance on a dataset derived from real-world change histories. Although \bdiff is text-based and language-independent, we aimed to create a dataset whose programming languages represent the major execution paradigms, as different languages exhibit distinct textual structures and formatting conventions. We therefore selected Java, Python, and XML, which represent compiled, interpreted, and declarative (markup) paradigms, respectively. For Java and Python, we used the \textit{GhPython} and \textit{GhJava} datasets~\cite{Falleri2024Fine}, which were constructed via stratified sampling to evaluate GumTree. Each dataset is built from the commit histories of 10 popular GitHub projects, containing approximately 100 file pairs (i.e., 100 cases) per project. For XML, we followed the same methodology to create a comparable dataset, named GhXML, from 10 popular GitHub projects that use XML. Consequently, we have a total of 2,977 cases with textual differences, containing 991, 999, and 987 cases for Python, Java, and XML, respectively.
\par For each case, we ran \bdiff and all baseline tools to compute the \textit{ES} on a desktop computer with a 3.6GHz Intel Xeon W-2223 CPU with 16GB of RAM. We set the analysis time limit at one hour. Consequently, among the 2,977 cases that have textual differences, Git diff successfully analyzed all the file pairs; ldiff and \bdiff could not complete the analysis process in the time limit for one file pair (scikit-learn/scikit-learn, cd076af, pendigits.py), which stores a large number of pure numbers; GumTree encountered 69 analysis errors on XML files and reported no differences in 128 cases, as it cannot detect formatting changes (e.g., the cases in Appendix~\ref{gt_no_format}); GPT-5-mini can successfully analyze 2,797 cases (the remaining 200 cases contain 183 no-difference cases and 17 failure cases); and Qwen3-32B can successfully analyze 2,443 cases (the remaining 554 cases contain 52 no-difference cases and 502 failure cases). To ensure a fair comparison, we retained 2,244 cases, where all tools successfully detected differences, as the dataset for our quantitative analysis.
\subsubsection{Base diff algorithm selection}
The \bdiff algorithm builds upon the deleted and added lines identified by a base diff algorithm. Consequently, different base diff algorithms can yield different results (e.g., different change hunks~\cite{Yusuf2019How}) for \bdiff. To evaluate this influence and select an optimal base algorithm, we computed the \textit{ES} for all file pairs in our dataset using \bdiff with each of Git's four diff algorithms: Myers, \textit{Minimal}, \textit{Patience}, and \textit{Histogram}. We used \textit{ES} size as the quality metric. Consequently, among the 2,976 cases, 144 exhibited discrepancies in \textit{ES} size across algorithms. In these 144 cases, \textit{Histogram} produced the smallest \textit{ES} most frequently (75 times); meanwhile, it also achieved the smallest average \textit{ES} size (16.28) and the shortest average running time (0.12s). Therefore, we selected \textit{Histogram} as default base diff algorithm for \bdiff. Our results align with findings in~\cite{Yusuf2019How} that \textit{Histogram} outperforms Git's other three algorithms.

\subsubsection{Prompt design}
\begin{figure}[!t]
    \centering
\includegraphics[width=0.9\linewidth]{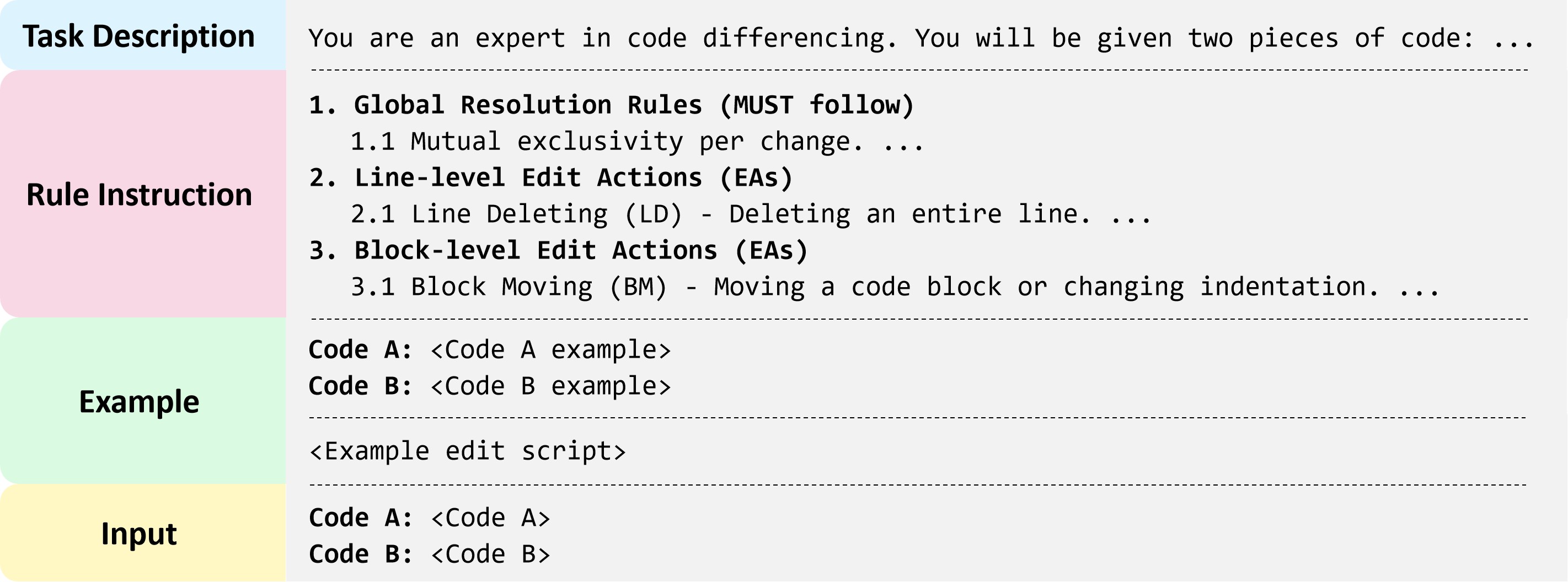}
    \caption{Overview of the prompt template.}
    \label{promptdesign_fig}
\end{figure}
We prompted the LLMs to replicate \bdiff's functionality and \textit{ES} format. Following established practices for instructing LLMs~\cite{ha2025one}, we employed a one-shot prompting strategy that integrates the task description, rule instructions, an example, and the input into a single unified prompt. Figure~\ref{promptdesign_fig} shows the prompt template. The \textit{Task Description} (in the system prompt) defines the LLM's role as a code differencing expert, tasked with analyzing \textit{Code A} (the left version) and \textit{Code B} (the right version) to output a JSON-like Python list representing the \textit{ES}. The \textit{Rule Instruction} specifies eight global resolution rules (e.g., minimizing \textit{ES} length, enforcing length constraints for the the blocks) and detailed formatting requirements for all the seven types of \textit{EA}s. The \textit{Example} section provides a real-world case with \textit{Code A}, \textit{Code B}, and the corresponding \textit{ES}, explicitly mapping code changes to compliant \textit{EA}s and illustrating how to apply the prompt's rules and formats. Finally, the \textit{Input} section uses a structured placeholder format that clearly defines the two code versions to be compared.

\subsection{Manual Evaluation}
To answer \textit{RQ2}, we consider the viewpoint of the developer. For her, what matters is that the computed \textit{ES} helps her efficiently understand the changes~\cite{Falleri2014Fine}. Accordingly, we designed a manual evaluation experiment to compare developers' perceptions of \bdiff's results against those of the baseline tools. We recruited 10 raters: 5 experienced industry software engineers and 5 graduate students majoring in Software Engineering. They evaluated a set of cases containing results from different differencing tools. To ensure representativeness, we used stratified sampling~\cite{Podgurski1999Estimation} to randomly select 100 Python, 100 Java, and 100 XML diff cases from the 2,997 cases in our dataset. For each language, 50 cases included block-level \textit{EA}s and 50 did not. For each case, we generated differencing results using \bdiff and all baseline tools. While \bdiff and GumTree natively provide web-based visualizations, other tools only output \textit{ES}s in text format. To standardize the evaluation interface, we developed an independent diff-result visualization tool\footnote{https://github.com/bdiff/Textual-Diff-Visualization-Exporter} that accepts \bdiff's \textit{ES} format as input and can export visualized differencing results as web pages. We then adapted the results of Git, LLMs, and ldiff\footnote{The line movements in ldiff are treated as single-line \textit{BM}s in \bdiff} to \bdiff's \textit{ES} format and obtained visual diff results of text-based and LLM-based differencing tools in a consistent style. This enables raters to compare all tools' outputs conveniently and intuitively while mitigating bias from inconsistent result presentation. Additionally, we provided commit messages for each case, which describe the change intent and facilitate meaningful assessment of differencing result quality.
\par The evaluation was conducted centrally in an online meeting. Raters were randomly divided into two groups (A and B) of five. Each group evaluated the complete case set, but each rater received a random subset of 60 cases (30 with block-level \textit{EA}s and 30 without). At the meeting, the first author of this paper introduced the experiment's background, objectives, and requirements, and emphasized that the results would be published to ensure accountability and objectivity. Raters then independently reviewed and scored the differencing results from all six tools for each case. Scores ranged from 1 to 6 (with higher values indicating better quality), representing a ranking of the tools' outputs; ties were allowed. Due to random selection, some cases might not have been successfully analyzed by certain tools (e.g., LLMs); for these, raters were instructed to assign a score of 1 to the corresponding tool. For each case, a comment field collected general feedback. For the 30 cases involving block-level \textit{EA}s, an additional field invited specific comments on the correctness of these block-level \textit{EA}s. After the evaluation, raters discussed the tools' performance and provided valuable suggestions for improving \bdiff, offering insights for our qualitative analysis.
\subsection{Mutation-based Evaluation}
To objectively assess the correctness of the generated \textit{ES} of \bdiff (\textit{RQ3}), we employed a mutation-based evaluation method~\cite{Roy2021The, 2009RoyDetection, LHDiffAsaduzzaman2013}. This method involves randomly injecting changes into a given code file and then analyzing how well \bdiff's computed \textit{ES} aligns with the ground-truth \textit{ES}. We implemented an algorithm that takes a file as input, randomly mutates it using the seven \textit{EA} types that \bdiff can identify (Section~\ref{AO}), and outputs the modified file along with the ground-truth \textit{ES}. To ensure the accuracy of the evaluation results, we configured the parameters of the mutation algorithm to be identical to those used by \bdiff.
\par Generally, the algorithm consists of the following three steps: 1) Data preparation. Using the input file (the left version), we create a list called \textit{right list} that contains a sequence of dictionaries, one for each line in the input file. This list represents the current state of the file during the mutation process and corresponds to the right version. The list's indices are used to determine the current line numbers, as these will dynamically change after \textit{EA}s are applied. In addition to the line content, each dictionary stores the corresponding line number from the left version, which is used to locate the correct position for subsequent mutations. 2) Code mutation. We first generate a random upper bound for the number of \textit{EA}s. The process then iteratively generates random changes in a top-to-bottom order, starting from the beginning of the file. In each iteration, we randomly select a position and an \textit{EA}, generate the corresponding change, and locate and modify the target line(s) in the \textit{right list}. The process terminates when the number of \textit{EA}s exceeds the maximum limit or the edit position moves beyond the end of the file. 3) \textit{ES} and right-version generation. Using the final \textit{right list}, we update the line numbers recorded in the \textit{EA}s and output the final \textit{ES} along with the right-version file.
\par We executed this algorithm to automatically mutate a random left version of the 2,997 cases in our dataset, obtaining the corresponding generated right versions and their ground-truth \textit{ES}s. We then ran \bdiff on these left/right version pairs to obtain the computed \textit{ES}s. Finally, we compared the \textit{EA}s in \bdiff's computed \textit{ES}s with those in the ground-truth \textit{ES}s. We define two \textit{ES}s as equivalent if they contain exactly the same \textit{EA}s. The order of \textit{EA}s within an \textit{ES} is not considered, as it is not unique (e.g., the order of two \textit{BC}s). To formalize our evaluation criteria, we define two \textit{EA}s as equivalent if they match in the following three aspects: (1) \textit{EA} type; (2) source line number(s) (applicable to non-\textit{LA} \textit{EA}s); and (3) target line number(s) (applicable to non-\textit{LD} \textit{EA}s). To ensure the reliability of our findings, we repeated the entire process for 3,000 mutations. We computed the matching rate, defined as the proportion of ground-truth \textit{EA}s that \bdiff can correctly identify, to evaluate \bdiff's correctness in \textit{ES} computation. Consequently, among the mutated 3,000 cases, the mean and median sizes of ground-truth \textit{ES} are 14.1 and 5.0, respectively, which are close to the mean and median (16.3 and 4.0) of the \textit{ES} sizes identified by \bdiff in real-world dataset.
\subsection{RQ1: Edit-script Size}
\begin{figure}[!t]
    \centering
    \includegraphics[width=0.8\linewidth]{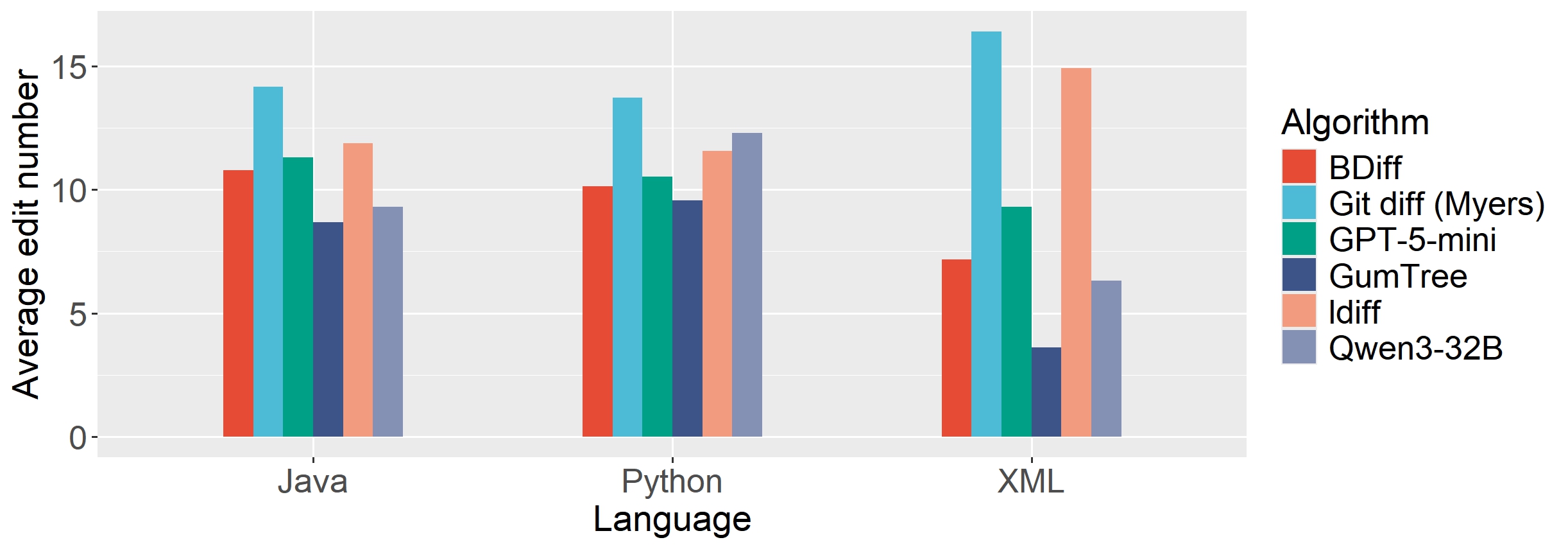}
    \caption{Average edit-script size of different languages}
    \label{nedits_avg}
\end{figure}
\begin{table}[!t]
\begin{threeparttable}
    \footnotesize
    \centering
    \renewcommand{\arraystretch}{1.1}
    \caption{Descriptive statistics of the \emph{ES} sizes of the differencing algorithms}
    \begin{tabular}{l l >{\raggedleft}p{1cm} >
    {\raggedleft}p{1cm} >
    {\raggedleft}p{1cm} >{\raggedleft}p{1cm} >{\raggedleft}p{1cm} m{1.7cm}}
\toprule
\specialrule{0em}{0.5pt}{0.5pt}
Approach type & Diff algorithm & 25\% & 50\% & Mean & 75\% & Max & Histogram\tnote{1} \\ \noalign{\smallskip}\hline\noalign{\smallskip}
 Tree-based & GumTree & 1.0 & 2.0 & 7.1 & 6.0 & 460.0 & \includegraphics[height=0.45cm]{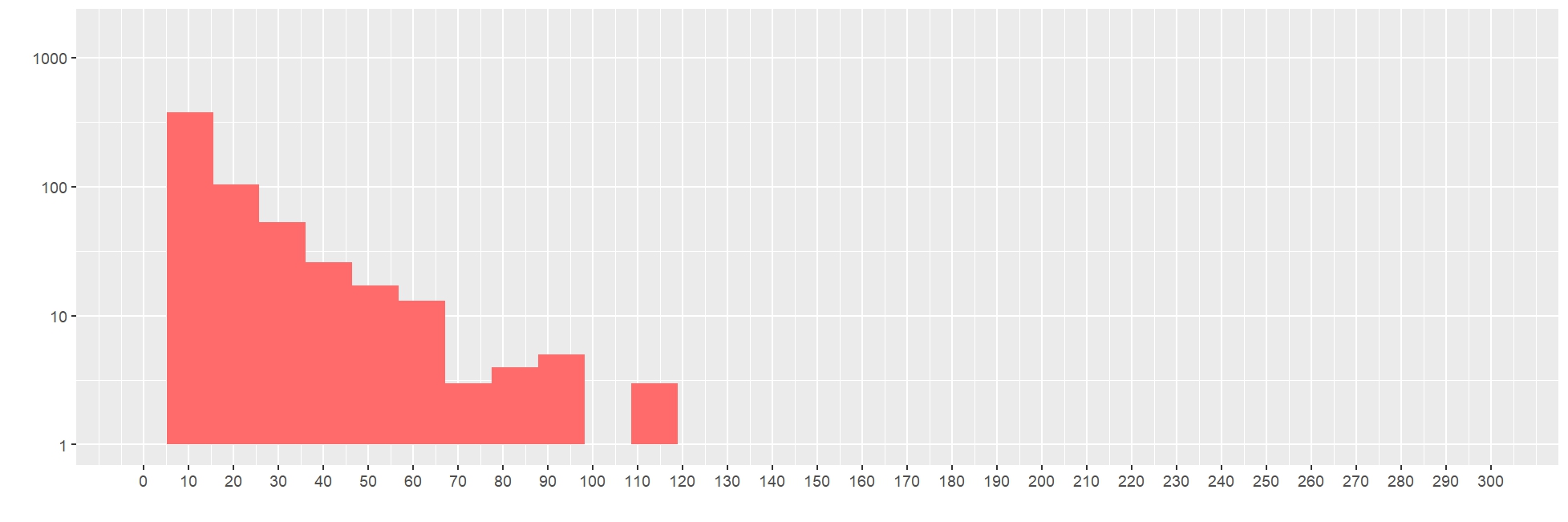} \\
 \hline
\multirow{3}{*}{LLM-based}    & GPT-5-mini & 1.0 & 3.0 & 10.3 & 10.0 & 490.0 & \includegraphics[height=0.45cm]{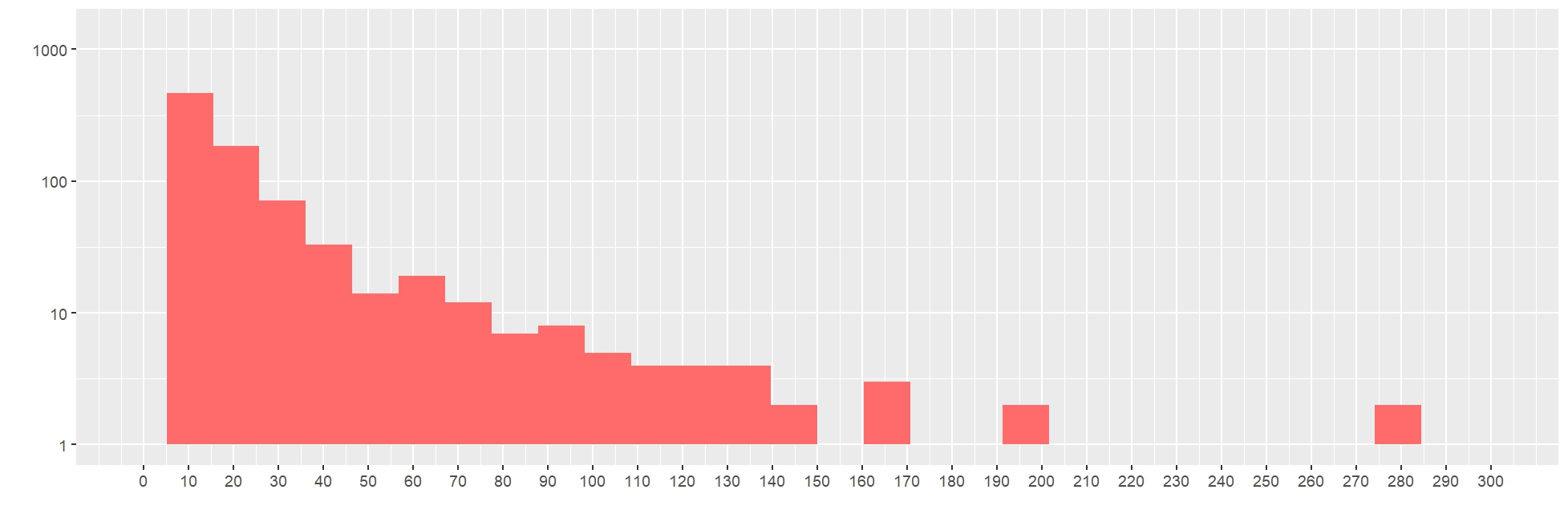} \\
    & Qwen3-32B & 1.0 & 2.0 & 9.1 & 7.0 & 893.0 & \includegraphics[height=0.45cm]{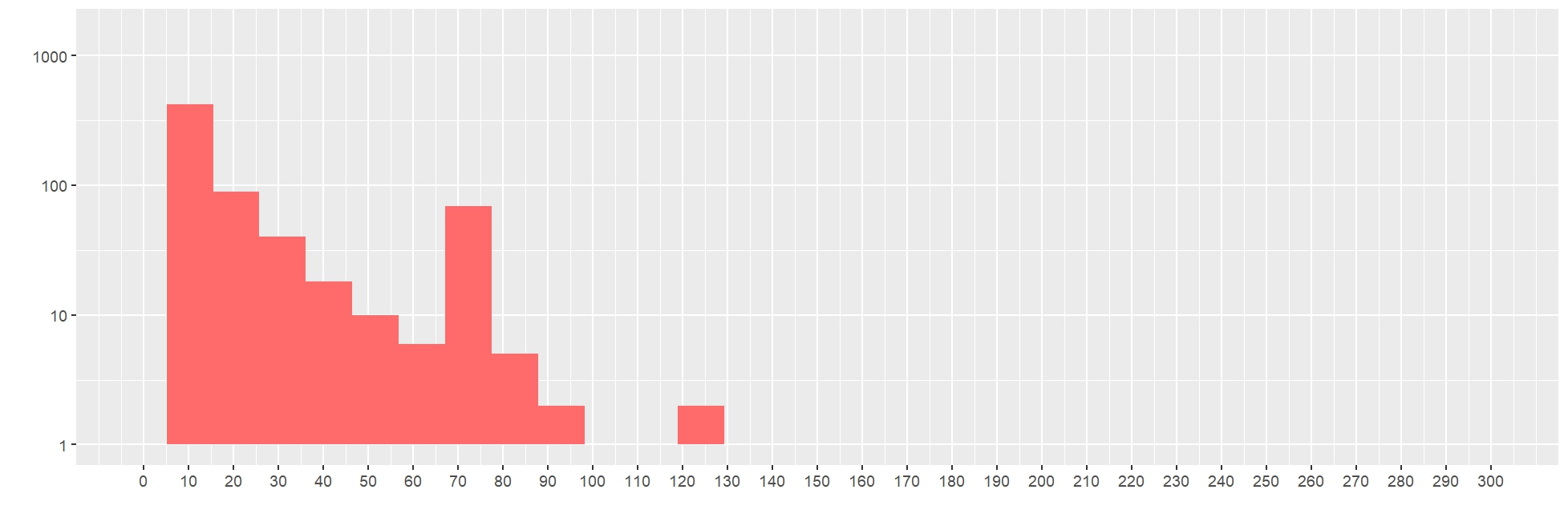} \\
    \hline
\multirow{4}{*}{Text-based} & Git diff (Myers) & 2.0 & 5.0 & 14.9 & 14.0 & 698.0 & \includegraphics[height=0.45cm]{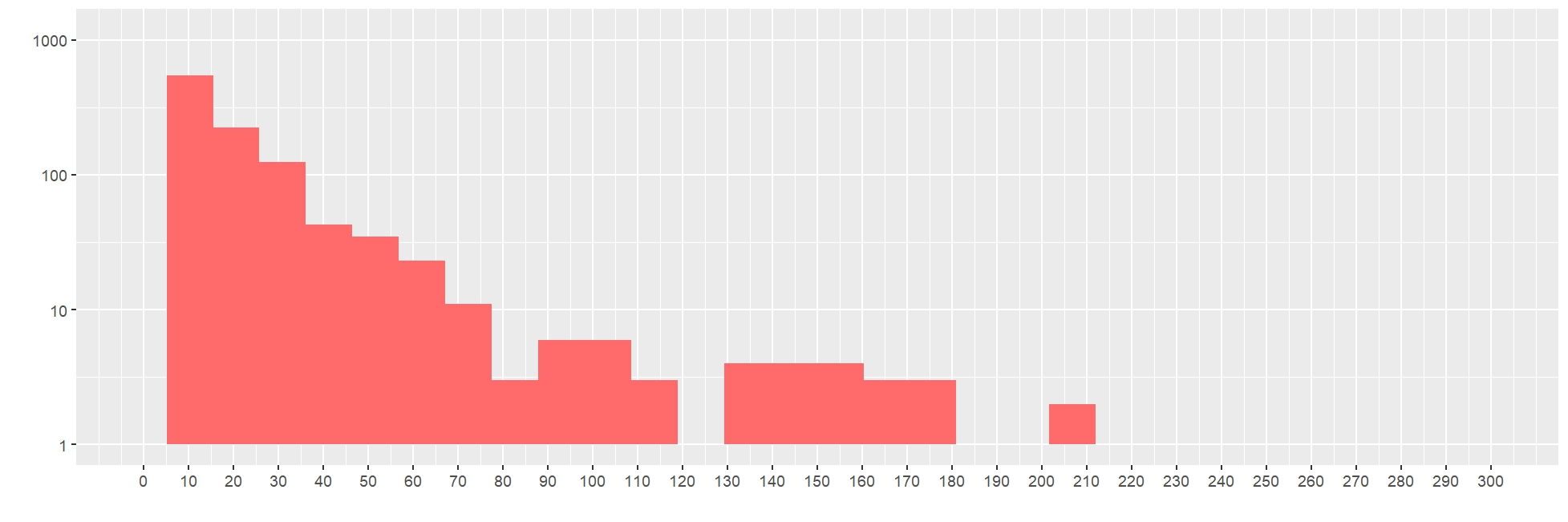} \\ 
    & ldiff & 1.0 & 4.0 & 12.9 & 11.0 & 698.0 & \includegraphics[height=0.45cm]{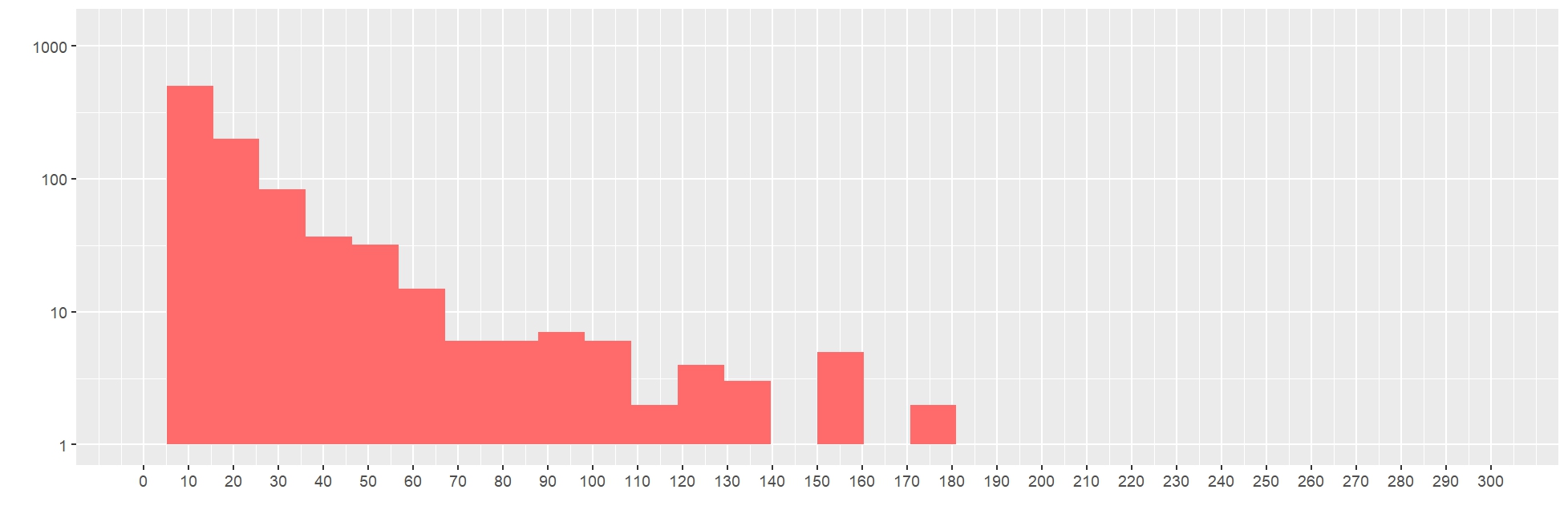} \\
    & \textbf{\bdiff} & \textbf{1.0} & \textbf{3.0} & \textbf{9.2} & \textbf{10.0} & \textbf{178.0} & \includegraphics[height=0.45cm]{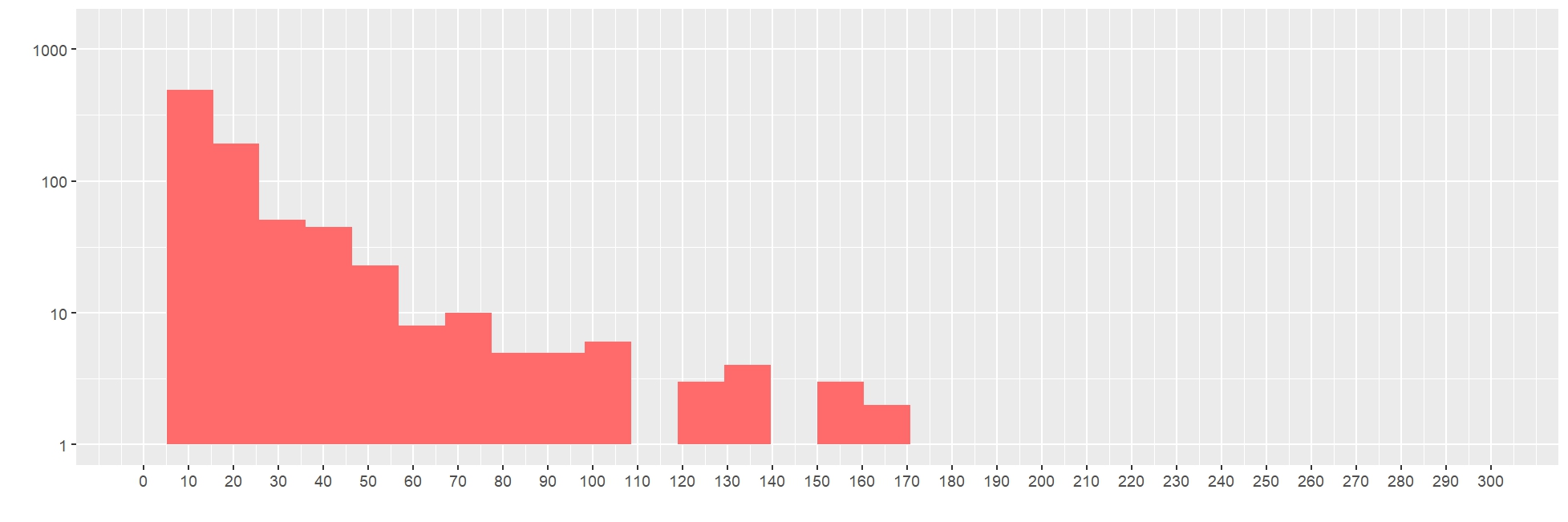} \\
     \bottomrule
    \end{tabular}
    \label{es_size}
    \begin{tablenotes}
    \scriptsize \item[1] Histograms are in log scale.
    \end{tablenotes}
    \end{threeparttable}
\end{table}
To evaluate \bdiff's performance in terms of \textit{ES} size, we first compared the descriptive statistics of the \textit{ES} sizes generated by \bdiff and the baseline approaches. The results are shown in Table~\ref{es_size}. Note that when calculating \textit{ES} size, we also include \textit{LU}s that occur within \textit{BM}s and \textit{BC}s.
In general, \bdiff generates the smallest \textit{ES} among the three text-based code differencing algorithms. The average \textit{ES} size of \bdiff is 37.9\% and 28.5\% lower than that of Git diff (Myers) and ldiff, respectively. In particular, for XML files, the average \textit{ES} size of \bdiff is 56.3\% and 52.0\% lower than that of Git diff and ldiff, respectively (see Figure~\ref{nedits_avg}). We attribute this to the fact that XML elements often share identical tags and similar structures, leading to \bdiff's \textit{EA}s that contain more \textit{BM}s and \textit{BC}s. In practice, developers frequently perform such edits when modifying XML files. Specifically, in 1,709 (76.2\%) cases, the \textit{ES} from \bdiff is shorter than that of Git diff; in 746 (33.2\%) cases it is shorter than that of ldiff; and in 1,345 (59.9\%) cases, \bdiff produces an \textit{ES} of the same size as ldiff. The average \textit{ES} size of GumTree is 22.8\% lower than that of \bdiff, while the \textit{ES}s generated by the two LLMs are generally comparable to those of \bdiff. We further used the paired nonparametric Wilcoxon signed-rank test to examine the significance of the differences in \textit{ES} size between \bdiff and each of the baseline tools. The results show that all differences are significant at the 0.0001 level. The effect size is small for Myers (Cliff's $\delta$: -0.196) and negligible for ldiff (Cliff's $\delta$: -0.045), GumTree (Cliff's $\delta$: 0.133), \textit{GPT-5-mini} (Cliff's $\delta$: 0.015), and \textit{Qwen3-32B} (Cliff's $\delta$: 0.110). 
\par We manually analyzed cases where the lengths of \textit{ES}s generated by \bdiff and GumTree significantly differ and found that GumTree's shorter \textit{ES}s are mainly attributable to two reasons: (1) GumTree identifies changes to multiple lines belonging to a single AST node as a single \textit{EA}, e.g., updating a multi-line comment (Figure~\ref{update_ast} in Appendix~\ref{ast_change}) and adding a complete function definition (Figure~\ref{insert_ast} in Appendix~\ref{ast_change}); and (2) GumTree is insensitive to code formatting changes, e.g., it does not identify \textit{LM}s (Figure~\ref{no_format1} in Appendix~\ref{gt_no_format}), \textit{LS}s, or changes in code block indentation (Figure~\ref{no_format2} in Appendix~\ref{gt_no_format}). Additionally, we observed that for certain \emph{BM}s identified by \bdiff, Gumtree would identify the lines within these \emph{BM}s as individual line movements when they are not treated as an integral AST node~(Appendix~\ref{gumtree_line_move_appendix}).
\begin{tcolorbox}[
    enhanced,
    colback=gray!16,
    boxrule=0pt,
    arc=3pt,
    left=6pt, right=6pt, top=6pt, bottom=6pt,
    boxsep=0pt,
    colupper=black]
\textbf{Summary}: 
In terms of \textit{ES} size, \bdiff outperforms text-based differencing tools, reducing the average \textit{ES} size by at least 28\%. It performs roughly on par with LLM-based tools, though its \textit{ES} is 23\% longer than that of the AST-based differencing tool GumTree.
\end{tcolorbox}

\subsection{RQ2: Result Quality of the Differencing Tools}
\label{res_quality}
To assess the reliability of the ratings, we computed the linearly weighted Cohen's Kappa ($\kappa$) coefficient to measure inter-rater agreement between the two groups of raters, who employed a 1–6 point scoring scale. The analysis yielded a weighted $\kappa$ of 0.61 with a $p$-value < 0.001, indicating statistically significant substantial agreement between the two rater groups~\cite{Landis1977TheMO, 2005Understanding}.
\begin{figure}
    \centering
    \includegraphics[width=1.02\linewidth]{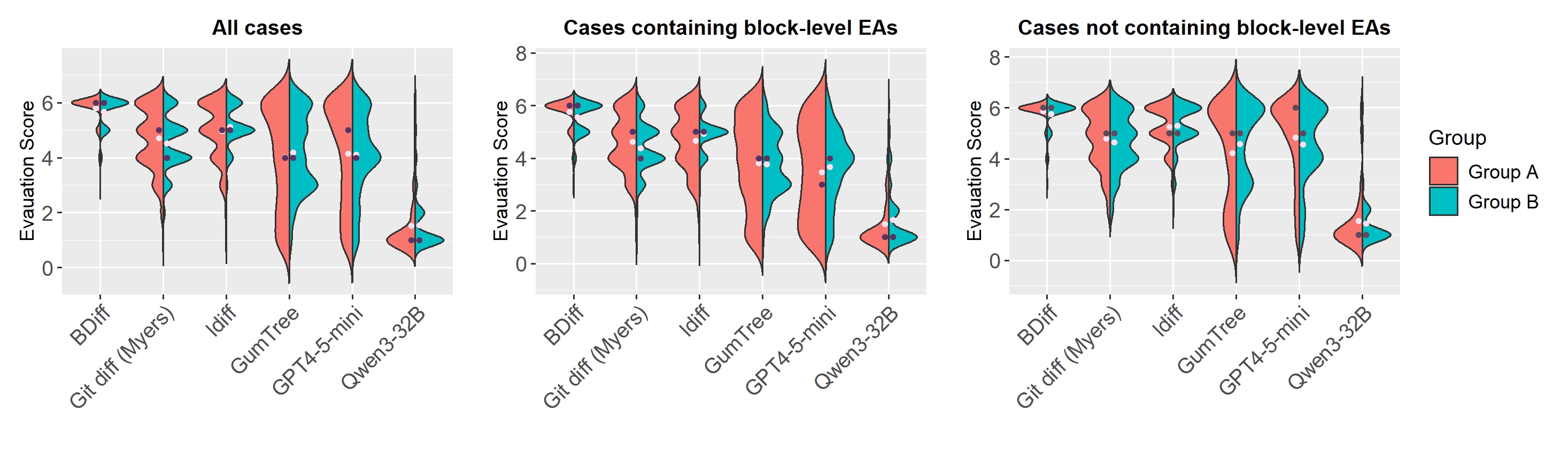}
    {\raggedright\footnotesize\textit{\textbf{Note}:} The points \begin{tikzpicture}\fill[mean_color] (0,0) circle (2pt); \end{tikzpicture} and \begin{tikzpicture}\fill[median_color] (0,0) circle (2pt); \end{tikzpicture} in the violin plots represent the mean and median, respectively.\par}
    \caption{Evaluation results of the result quality of the tools}
    \label{manual_eval_fig}
\end{figure}
We first analyzed the raters' scores on the differencing results from the manual evaluation experiment (see Figure~\ref{manual_eval_fig}). Overall, the average scores given by the 10 raters to the results generated by \bdiff, ldiff, Git, GPT-5-mini, GumTree, and Qwen3-32B are 5.73, 5.04, 4.61, 4.13, 4.09, and 1.53, respectively. Among \bdiff's 300 results, 63.3\% received the highest score of 6 from both raters, and only 5 cases (1.7\%) received a score below 5 from both. During the discussion segment, raters generally agreed that \bdiff's results were intuitive and concise. In particular, raters reported that the block-level \emph{EA}s identified by \bdiff are of high readability: ``\emph{\bdiff's BCs are highly intuitive}'' [A5] and ``\textit{\bdiff performs best in text block handling and can recognize a variety of complex scenarios}'' [B5]. We also found that for a considerable number of cases with block-level \emph{EA}s, \bdiff's results can intuitively reflect developers' actual editing intentions and facilitate readers in efficiently understanding the changes, e.g., \dashuline{\textit{reordering of assignment statement blocks}} (Appendix~\ref{reorder_assignment_appendix}), \dashuline{\textit{indenting a code block for conditional branching}} (Appendix~\ref{Intenting_block_appendix}), \dashuline{\textit{reusing an XML block or a function}} \dashuline{\textit{ definition}} (Appendix~\ref{Reusing_XML_appendix},~\ref{Reusing_func_appendix},~\ref{Reusing_xml_appendix2}), and \dashuline{\textit{reformatting code}} (Appendix~\ref{Splitting_appendix2}). For all tools except for Qwen3-32B, the average scores of the cases that contain block-level \emph{EA}s (i.e., the \textit{block cases}) were lower than those of the cases not containing block-level \emph{EA}s (i.e., the \textit{non-block cases}): 5.65 \textit{vs} 5.81, 4.79 \textit{vs} 5.27, 4.5 \textit{vs} 4.72, 3.56 \textit{vs} 4.70, and 3.80 \textit{vs} 4.40, and 1.57 vs 1.49 for \bdiff, ldiff, Git, GPT-5-mini, GumTree, and Qwen3-32B, respectively. We believe this is mainly due to the complexity of the block cases: the average \emph{ES} sizes (computed by Git diff) of the block cases and non-block cases are 75.0 and 11.1, respectively. Raters also reported that in some complex change scenarios, the identified block-level \textit{EA}s are inaccurate and can hinder comprehension, e.g., ``\textit{There are too many BCs and BMs, which is confusing and hard to understand}'' [A5], and ``\textit{Various EAs are intertwined, making it rather chaotic. It's not as easy to understand at a glance as Git}'' [A2, A3, B1].
\par Due to its ability to identify line changes, ldiff received higher average scores than Git. Raters reported that ``\textit{Git only identifies \textit{LD}s and \textit{LA}s, which makes it difficult to understand in scenarios involving a large number of consecutive \textit{EA}s}''~[A4]. Notably, for the same case, ldiff may split a single \textit{BM} (as identified by \bdiff) into multiple line moves, resulting in ``\textit{Longer ESs that are neither intuitive nor easy to understand}" [A3].
\begin{figure}
    \centering
    \includegraphics[width=\linewidth]{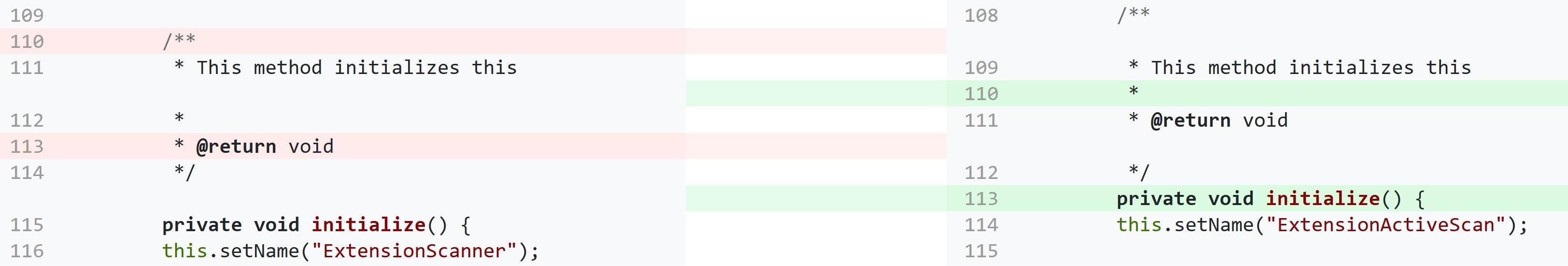}
    \caption{A diff snippet generated from GPT-5-mini}
    \label{gpt-error}
\end{figure}
\par GumTree generally received lower scores compared to text-based differencing tools. Our results on GumTree's performance relative to Git diff are lower than those reported in a prior study~\cite{Falleri2014Fine}.
We attribute this discrepancy to two possible factors: (1) the cases in the previous experiment contained fewer changes, while our experiment included cases with complex changes; and (2) prior raters were all from research teams, while half of our raters are industry engineers who are relatively unfamiliar with AST-based differencing results, despite relevant introductions before the evaluation. In our experiment, three-fifths of the raters from industry reported that GumTree's results were difficult to understand, e.g., ``\textit{It is challenging to understand Gumtree's results}''~[A3], and ``\emph{Gumtree relies on manual correlation for side-by-side line comparison, which is user-unfriendly and lacks proper alignment}''~[A5].
\par Finally, regarding the two LLM-based tools, the average score of the open-source model Qwen3-32B (1.53) is much lower than that of GPT-5-mini (4.13), and three raters felt that Qwen3-32B delivered the worst performance [A3, B4, B5]. The raters reported that both GPT-5-mini and Qwen3-32B exhibited numerous recognition errors and significant hallucinations in the diff generation task, e.g., ``missing certain EAs'' [B2], ``incorrectly identifying line numbers'' [B2], and ``incorrectly identifying EAs'' [A5]. Even in simple diff scenarios, these LLMs can introduce naive errors. As shown in Figure~\ref{gpt-error}, GPT-5-mini cannot correctly identify simple \textit{LD}s and \textit{LA}s.
\begin{tcolorbox}[
    enhanced,
    colback=gray!16,
    boxrule=0pt,
    arc=3pt,
    left=6pt, right=6pt, top=6pt, bottom=6pt,
    boxsep=0pt,
    colupper=black]
\textbf{Summary}:
\bdiff achieved the highest average score, primarily due to the superior readability of its differencing results. In contrast, the outputs from LLM-based tools received relatively low scores and were plagued by hallucinations.
\end{tcolorbox}

\subsection{RQ3: ES Correctness of \bdiff}
\begin{table}[t]
\scriptsize
\renewcommand{\arraystretch}{1.4}
\centering
\caption{Experimental results of the mutation-based evaluation} 
\label{mutation_eval}
\begin{tabular}{lccccclcccccccc}
\toprule
\multirow{2}*{Language} & \multirow{2}*{\# Files} & \multicolumn{2}{c}{Ground-truth ES size} & \multicolumn{2}{c}{\bdiff ES size} & \multicolumn{8}{c}{Avg. matching rate} \\ 
\cline{3-4} \cline{5-6} \cline{7-14}
& & Mean & Median & Mean & Median & Total & LD & LA & LU & LS & LM & BM & BC\\
\hline
Java  & 985 & 16.4 & 5.0 & 19.8 & 5.0 & 0.948 & 0.968 & \cellcolor{light_green} 0.995 & 0.914 & \cellcolor{light_green} 0.997 & 0.955 & \cellcolor{light_red} 0.810 & \cellcolor{light_red} 0.808\\
Python  & 1,015 & 17.0 & 5.0 & 19.4 & 5.0 & 0.953 & 0.948 & \cellcolor{light_green} 0.995 & 0.948 & \cellcolor{light_green} 0.986 & 0.937 & \cellcolor{light_red} 0.832 & \cellcolor{light_red} 0.824\\
XML  & 1,000 & 8.7 & 4.0 & 9.5 & 4.0 & 0.957 & 0.967 & \cellcolor{light_green} 0.999 & 0.962 & \cellcolor{light_green} 0.992 & 0.967 & \cellcolor{light_red} 0.834 & \cellcolor{light_red} 0.887\\
\hdashline[3pt/2pt]
\textbf{Total}  & 3,000 & 14.1 & 5.0 & 16.2 & 5.0 & 0.953 & 0.961 & \cellcolor{light_green} 0.996 & 0.940 & \cellcolor{light_green} 0.991 & 0.952 & \cellcolor{light_red} 0.825 & \cellcolor{light_red} 0.828\\
\toprule
\end{tabular}
\end{table}
Table~\ref{mutation_eval} presents the results of the mutation-based evaluation. Overall, no significant differences in ground-truth \textit{ES} size are observed across languages. The mean and median sizes of ground-truth \textit{ES} are 14.1 and 5.0, respectively. By comparison, the average \textit{ES} size computed by \bdiff across the mutation cases is slightly larger (16.2). Among the 3,000 mutations, the average \textit{ES} matching rate reaches 95.3\%, and 2,473 (82.4\%) cases achieve a 100\% matching rate, demonstrating high identification accuracy of \bdiff. Among the seven \textit{EA} types, \textit{LA} and \textit{LS} achieve the highest matching rates (>99.0\%), while the block-level \textit{EA}s, \textit{BM} and \textit{BC}, yield relatively lower rates (82.5\% and 83.8\%, respectively). We therefore conducted a focused analysis of \bdiff's accuracy on \textit{BM} and \textit{BC}. Across all mutation cases, 1,075 cases generated 1,228 \textit{BM}s, and 1,167 cases generated 1,543 \textit{BC}s. \bdiff correctly identified 1,012 (82.4\%) of the \textit{BM}s and 1,278 (82.8\%) of the \textit{BC}s. We analyzed the misidentified cases and identified two primary causes: (1) inaccuracies in the Git diff output, such as misidentifying deleted and added lines, which propagates to \bdiff's identification of \textit{BM}s and \textit{BC}s (and can also lead to misidentifying of other \textit{EA}s such as \textit{LU}); and 2) the presence of multiple similar code blocks within the same file, where \bdiff may fail to select the correct block mapping during the optimal matching step~\circled{6}. 
\begin{tcolorbox}[
    enhanced,
    colback=gray!16,
    boxrule=0pt,
    arc=3pt,
    left=6pt, right=6pt, top=6pt, bottom=6pt,
    boxsep=0pt,
    colupper=black]
\textbf{Summary}: 
\bdiff achieves an accuracy of 95.3\% in identifying ground-truth \textit{EA}s, with fully correct \textit{ES} identification in 82.4\% of cases. Among all \textit{EA} types, the accuracy for block-level \textit{EA}s is relatively lower (approximately 83\%), primarily due to inaccuracies in the base diff algorithm's output and the presence of multiple similar code fragments within the same file.
\end{tcolorbox}
\subsection{RQ4: Runtime Performance}
We now analyze the runtime performance of \bdiff and the baseline tools on the 2,244 cases. For each case, we ran 10 times and retained the median value. For all tools, we only recorded the time taken to compute the \textit{ES}, excluding the time spent on visualizing differencing results. For GumTree, for instance, we used the ``\textit{TextDiff}'' mode to obtain the computed \textit{ES} in text format. Overall, the tools show no significant difference in average running time across different programming languages (see Figure~\ref{runningtime_avg}).
\begin{figure}
    \centering
\includegraphics[width=0.8\linewidth]{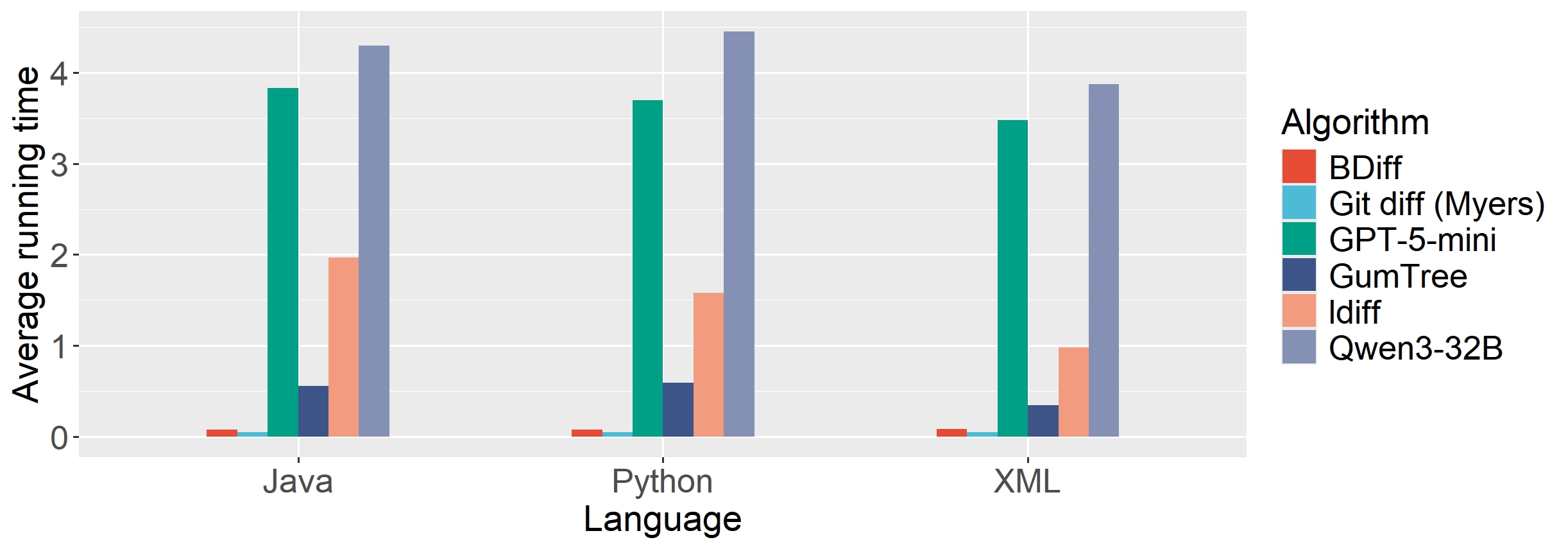}
    \caption{Average running time of different languages}
    \label{runningtime_avg}
\end{figure}
\begin{table}[!t]
\begin{threeparttable}
    \scriptsize
    \centering
    \renewcommand{\arraystretch}{1.2}
    \caption{Descriptive statistics of the runtimes (in seconds) of the diff algorithms}
    \label{runtimetable}
    \begin{tabular}{l l >{\raggedleft}p{0.9cm} >
    {\raggedleft}p{0.9cm} >
    {\raggedleft}p{0.9cm} >{\raggedleft}p{0.9cm} >{\raggedleft}p{0.9cm} >{\raggedleft}p{0.9cm} m{1.7cm}}
\toprule
\specialrule{0em}{0.5pt}{0.5pt}
Approach type & Diff algorithm & 25\% & 50\% & Mean & 75\% & Max & SD & Histogram\tnote{1} \\      		\noalign{\smallskip}\hline\noalign{\smallskip}
   Tree-based & GumTree & 0.426 & 0.637 & 0.643 & 0.764 & 1.950 & 0.256 &\includegraphics[height=0.45cm]{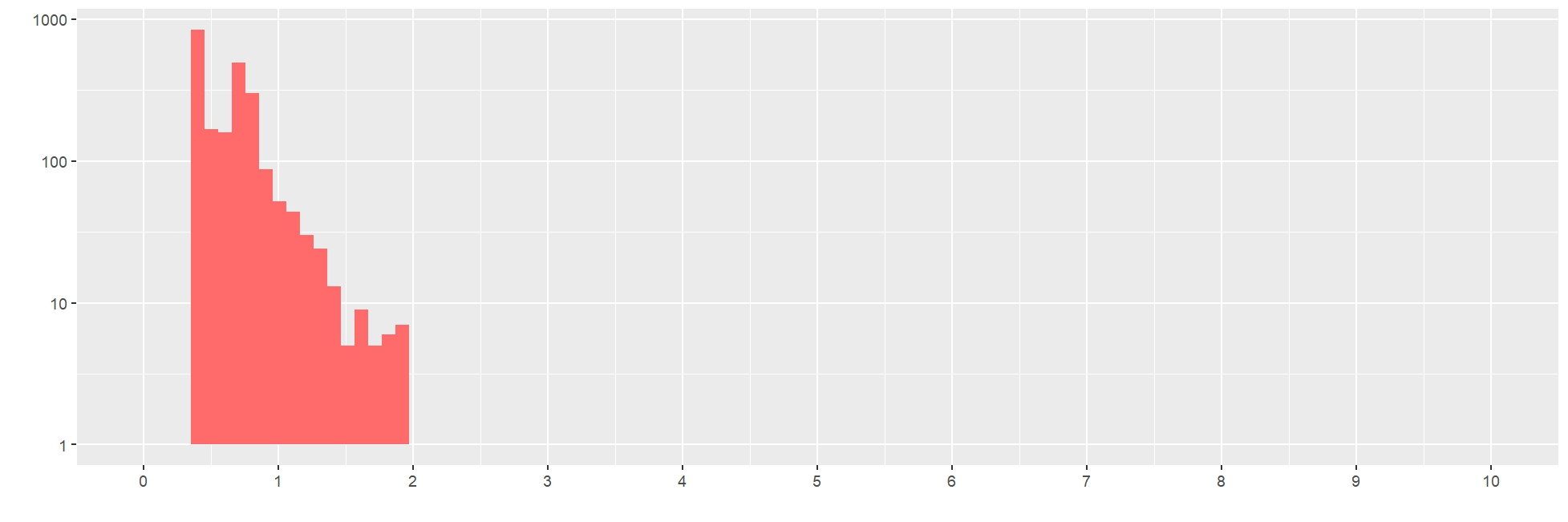} \\
   \hline
   \multirow{3}{*}{LLM-based} & GPT-5-mini & 18.473 & 29.897 & 38.076 & 48.723 & 874.001 & 39.633 & \includegraphics[height=0.45cm]{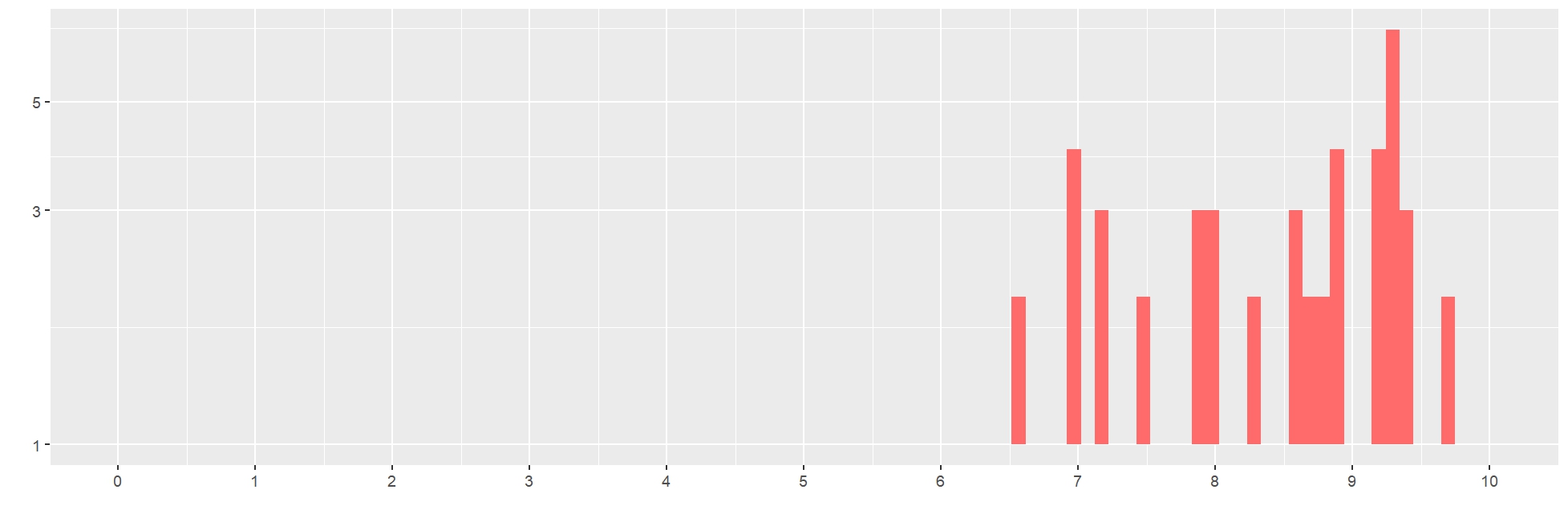} \\
     & Qwen3-32B & 16.413 & 37.398 & 66.681 & 89.805 & 1480.290 & 98.497&\includegraphics[height=0.45cm]{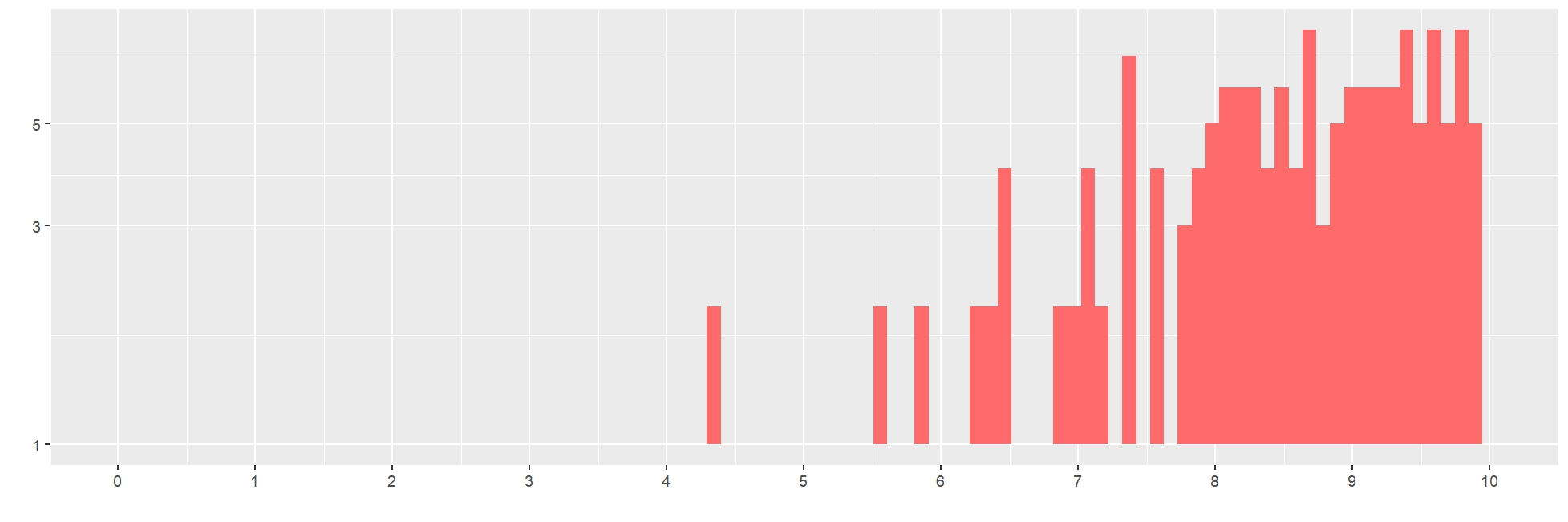} \\
     \hline
    \multirow{4}{*}{Text-based} & Git diff (Myers) & 0.050 & 0.051 & 0.051 & 0.052 & 0.057 & 0.002 &\includegraphics[height=0.45cm]{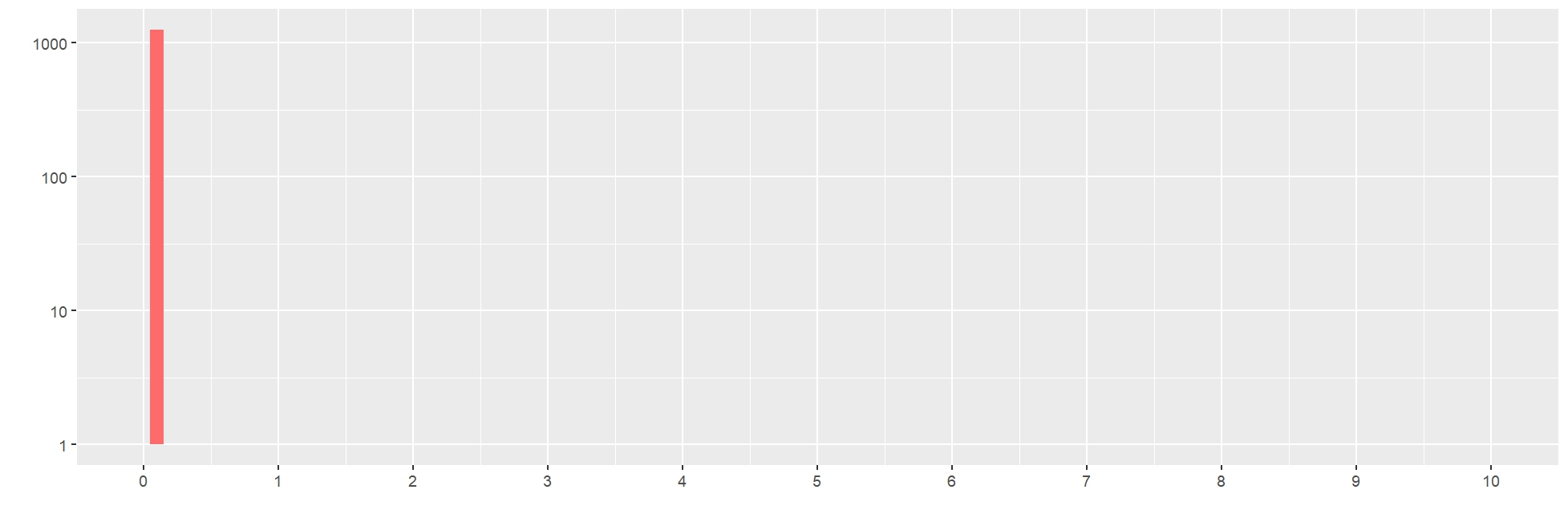} \\ 
    & ldiff & 0.467 & 0.547 & 3.766 & 1.472 & 946.784 & 27.581 & \includegraphics[height=0.45cm]{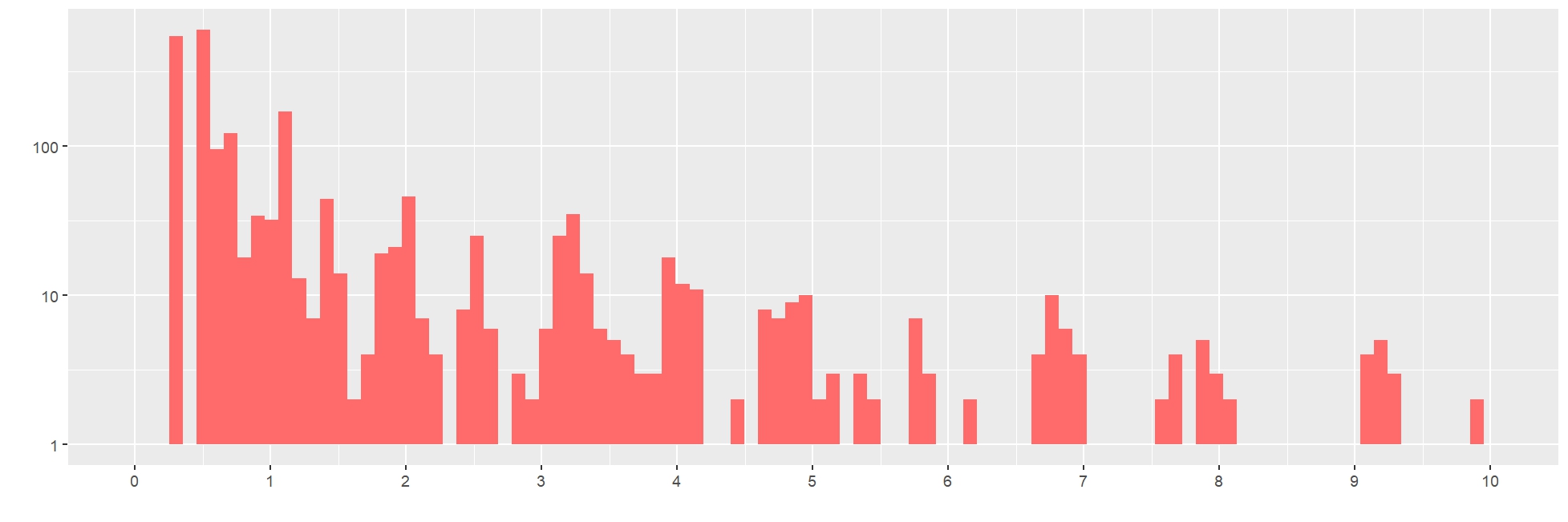} \\
    & \textbf{\bdiff} & \textbf{0.075} & \textbf{0.077} & \textbf{0.085} & \textbf{0.080} & \textbf{2.633} & \textbf{0.084} & \includegraphics[height=0.45cm]{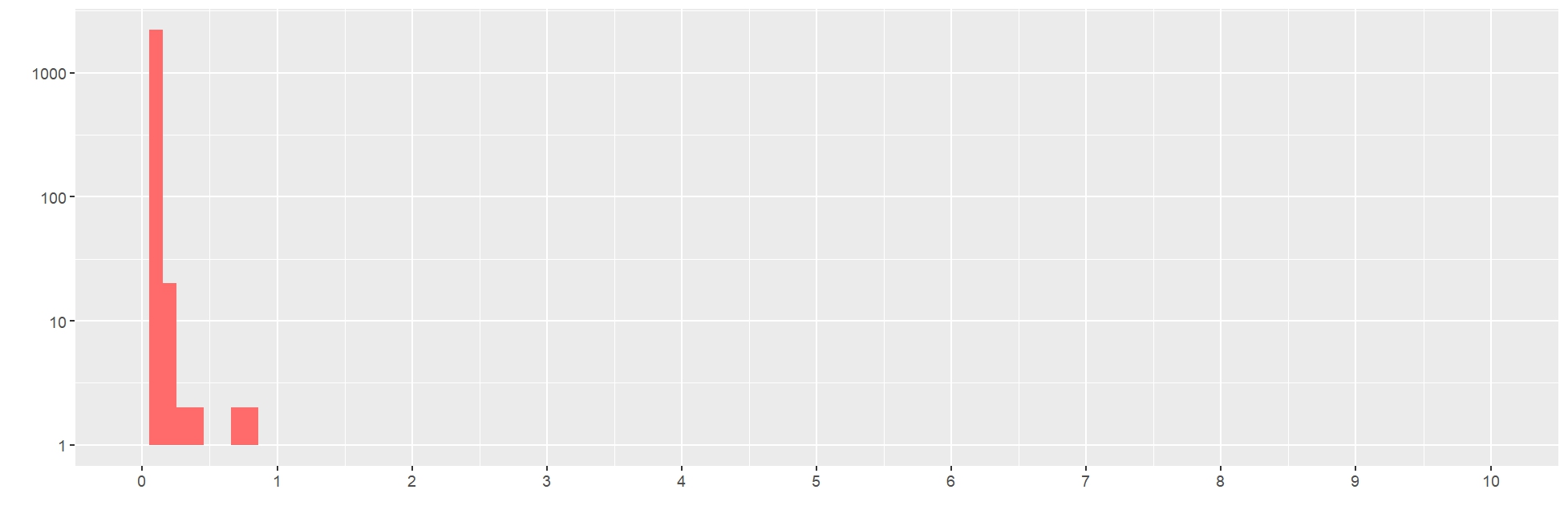} \\
     \bottomrule
    \end{tabular}
    \begin{tablenotes}
    \scriptsize \item[1] Histograms are in log scale.
    \end{tablenotes}
    \end{threeparttable}
\end{table}
The descriptive statistics are summarized in Table~\ref{runtimetable}. As expected, Git diff is the fastest due to its simpler identification capability, with an average running time of 0.051s. \bdiff ranks second, with an average running time of approximately 0.085s. \bdiff also exhibits good runtime stability: its standard deviation of 0.084s is the second lowest, just above Git's. The maximum runtime across our cases is 2.633s, which is completely acceptable in practice. The average running time of ldiff is 3.766s, but its running time is more variable, with a standard deviation of 27.581s. In our cases, 136 (6.1\%) take longer than 10 seconds, which may hinder interactive use by developers~\cite{Falleri2014Fine}. By contrast, the two LLM-based tools perform extremely poorly in terms of runtime: the average running times of GPT-5-mini and Qwen3-32B are 38.076s and 66.681s, respectively. Furthermore, only 72 cases (3.2\%) for GPT-5-mini and 170 cases (7.6\%) for Qwen3-32B complete within 10 seconds, making them impractical for interactive use.
\begin{tcolorbox}[
    enhanced,
    colback=gray!16,
    boxrule=0pt,
    arc=3pt,
    left=6pt, right=6pt, top=6pt, bottom=6pt,
    boxsep=0pt,
    colupper=black]
\textbf{Summary}: 
\bdiff exhibits short and stable runtime, with an average running time of less than 0.10 seconds and a standard deviation of around 0.08 seconds—ranking second only to Git. In contrast, the LLM-based differencing tools have average running times exceeding 30 seconds, making them impractical for interactive use.
\end{tcolorbox}

\section{Discussion}
\subsection{For Developers}
\subsubsection{Customizing \bdiff for Specific Scenarios}
\label{impli_dev1}
In our evaluation experiment, we have observed the phenomenon of ``\emph{one size does not fit all}'': the results of \bdiff generally received high ratings; however, for certain cases, \bdiff exhibits suboptimal performance under the same algorithmic settings. We believe that this can be primarily attributed to the inherent inaccuracies of the mapping step in a difference computing process, in which edit process information is missing and the left version and right version may have multiple same/similar lines/fragments. In response, developers can adjust the algorithm settings for specific scenarios to obtain better diff results. For example, developers can disable the identification of line updates within block-level \textit{EA}s, or set a larger minimum block size for \textit{BM}s and \textit{BC}s to reduce ``overly broad'' matches. Furthermore, developers can configure which \textit{EA} types to recognize. For complex cases, they may disable specific \textit{EA}s to simplify results to only include \textit{LD}s and \textit{LA}s, thereby improving results readability.

\subsubsection{Using BMs and BCs to support SE activities}
The ability to identify block-level \textit{EA}s is a key feature of \bdiff. Specifically, a \textit{BM} associates consecutive added lines with consecutive deleted lines, while a \textit{BC} associates consecutive added lines with consecutive lines in the left version. These block-level associations operate at a coarser granularity than the line-based associations used in traditional source tracking techniques~\cite{LHDiffAsaduzzaman2013}. Developers can leverage these associations to support several practical SE activities:
(1) Defect localization. By tracking code blocks through \textit{BM}s and \textit{BC}s, developers can more effectively pinpoint the source of defects. In particular, \textit{BC}s can help discover additional code blocks that may contain defects by revealing their relationship to existing code.
(2) Change analysis. Changes in \textit{BM} block positions, particularly those indicated by indentation changes, can be used to analyze modifications in the scope of related variables. Furthermore, \textit{BC}s can reveal code clones, suggesting potential refactoring opportunities such as abstracting common code blocks into shared functions or classes.

\subsection{For Researchers}
\subsubsection{\bdiff's support for downstream SE tasks}
Code differencing is a foundational technique in software maintenance and evolution research. To the best of our knowledge, \bdiff is the first to propose and implement block-level \textit{EA} recognition, with comprehensive \textit{EA} identification capabilities. This advance can significantly benefit many downstream SE research efforts based on differencing technology. We outline four main research directions: (1) Conflict detection. Current approaches primarily include text-based, AST-based, and semistructured methods. \bdiff can provide more block-aware, accurate text-based differencing results, which can be used to improve detection accuracy and reduce false positives. (2) Change pattern mining. With \bdiff's more accurate differencing results that align with developers' intended \textit{EA}s, researchers can conduct more precise studies on defect repair patterns and software evolution patterns. (3) Change summary generation. Researchers can generate more accurate change summaries~\cite{li2024only} from \bdiff's results. For example, changes involving only \textit{BM}, \textit{LS}, and \textit{LM} can produce refactoring-oriented summaries. (4) Contribution measurement. Our experimental results show that \bdiff reduces the average \textit{ES} size by 28\% compared with existing text-based differencing tools. \bdiff's results enable more accurate measurement of developers' code contributions. For instance, calculations involving \textit{BM} more accurately reflect actual contributions compared to traditional metrics based solely on \textit{LD}s and \textit{LA}s.

\subsubsection{Intelligent code differencing based on \bdiff} 
Existing code differencing techniques employ model-driven approaches that formalize problem-solving processes through explicitly defined mathematical or logical frameworks, enabling deterministic optimization. While these methods are static and efficient, they struggle to consistently achieve optimal performance across all scenarios, i.e., the ``one size does not fit all'' phenomenon noted in Section~\ref{impli_dev1}. To address this limitation, researchers could explore integrating data-driven methods (e.g., deep learning and reinforcement learning) with \bdiff's approaches and results to intelligently generate more accurate and contextually appropriate outcomes. Furthermore, our design of the weight calculation formula for the KM process is primarily based on empirical insights, lacking rigorous validation. For future work, researchers could leverage data-driven approaches to train more accurate and robust parameters.

\section{Threats to Validity}
\textbf{Construct validity}.
In this paper, we use \textit{ES} size as a metric to measure the quality of differencing results. However, a shorter \textit{ES} may be more difficult to understand than a longer one~\cite{Falleri2014Fine} and may even be incorrect. To complement this metric, we conducted both a qualitative manual experiment and a quantitative mutation-based evaluation. In the mutation-based evaluation, we use the matching rate as a measure of the correctness of \textit{ES} identification. This metric considers only the proportion of matched \textit{EA}s among the ground-truth \textit{EA}s, without accounting for over-identified \textit{EA}s. However, this limitation had minimal impact in our experiments because 82.4\% of the cases achieved perfectly matched \textit{ES}s, and the average \textit{ES} size of \bdiff (16.2) is slightly larger than that of the actual \textit{ES} (14.1).
\par \textbf{Internal validity}. In our experiments, we used \bdiff's default settings, which incorporate multiple parameter configurations such as a maximum line count of 8 for \textit{LS} and \textit{LM}, and a minimum block size of 2 for \textit{BM} and \textit{BC} (excluding lines containing only stop words). These defaults have not been validated through dedicated experiments to confirm they yield optimal differencing results. To mitigate this threat, we referenced default parameters from prior tools (e.g., the maximum line count for \textit{LS} and \textit{LM} in LHDiff) and randomly selected a subset of results for manual analysis to verify parameter effectiveness.
Another threat in the manual evaluation is the subjectivity of rater assessments. Raters with diverse professional backgrounds and experiences may have scored cases based on their prior familiarity with differencing tools, e.g., engineers who frequently use Git diff may be relatively unfamiliar with GumTree's AST-based approach, while graduate students often have research experience with GumTree. To address this, we: (1) introduced all tools during the evaluation briefing and analyzed examples of high-quality differencing results; (2) provided real-time clarification of raters' questions during evaluation; and (3) held post-evaluation discussions allowing raters to revise their scores before final submission. A third threat involves GumTree's XML backend, which is known to have issues. In our experiments, some XML cases triggered parsing errors during GumTree analysis. We mitigated this by excluding error cases from the dataset and validating the correctness of remaining results through random sampling.
\par \textbf{External validity}. Our dataset comprises 2,244 cases across three languages: Java, Python, and XML. However, we cannot claim these changes are representative of all changes in these languages~\cite{Falleri2024Fine}. Moreover, while \bdiff is language-independent, we cannot guarantee our results will generalize to other languages.

\section{Conclusion}
We have developed \bdiff, a novel text-based and block-aware approach for accurate code differencing. \bdiff builds upon the results of traditional differencing algorithms and uses the KM algorithm to compute an optimal \textit{ES}. We conducted a comprehensive evaluation, and the results demonstrate the effectiveness and practical utility of our approach. Our algorithm is implemented in an open-source, web-based visual differencing tool. We believe \bdiff can advance both software engineering practice and related research.

\begin{acks}
We would like to thank the experiment raters for their
time. We gratefully acknowledge the support from the National
Natural Science Foundation of China (Grant No.62302515). 
\end{acks}

\bibliographystyle{ACM-Reference-Format}
\bibliography{diff}

\newpage
\appendix

\section{Typical \bdiff Results with Block-level EAs}
\subsection{Changing the Orders of Initialization Assignment of Class Attributes}
\label{reorder_assignment_appendix}
\begin{figure}[H]
    \centering
    \includegraphics[width=\linewidth]{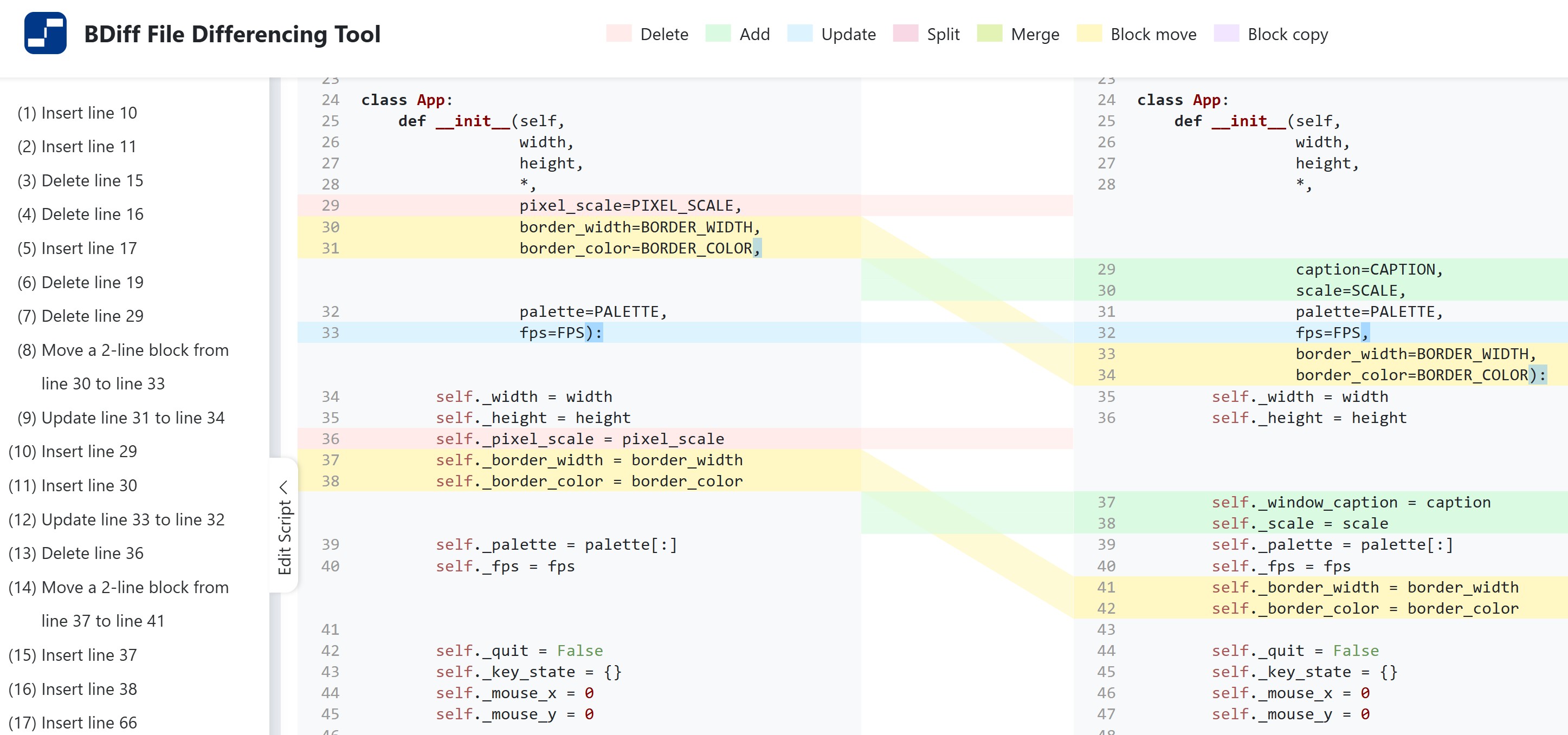}
    \caption{GitHub Project: kitao/pyxel. Commit: 3861523. File: app.py}
    \label{reorder_attr}
\end{figure}

\subsection{Indenting a Block for Conditional Branching}
\label{Intenting_block_appendix}
\begin{figure}[H]
    \centering
    \includegraphics[width=\linewidth]{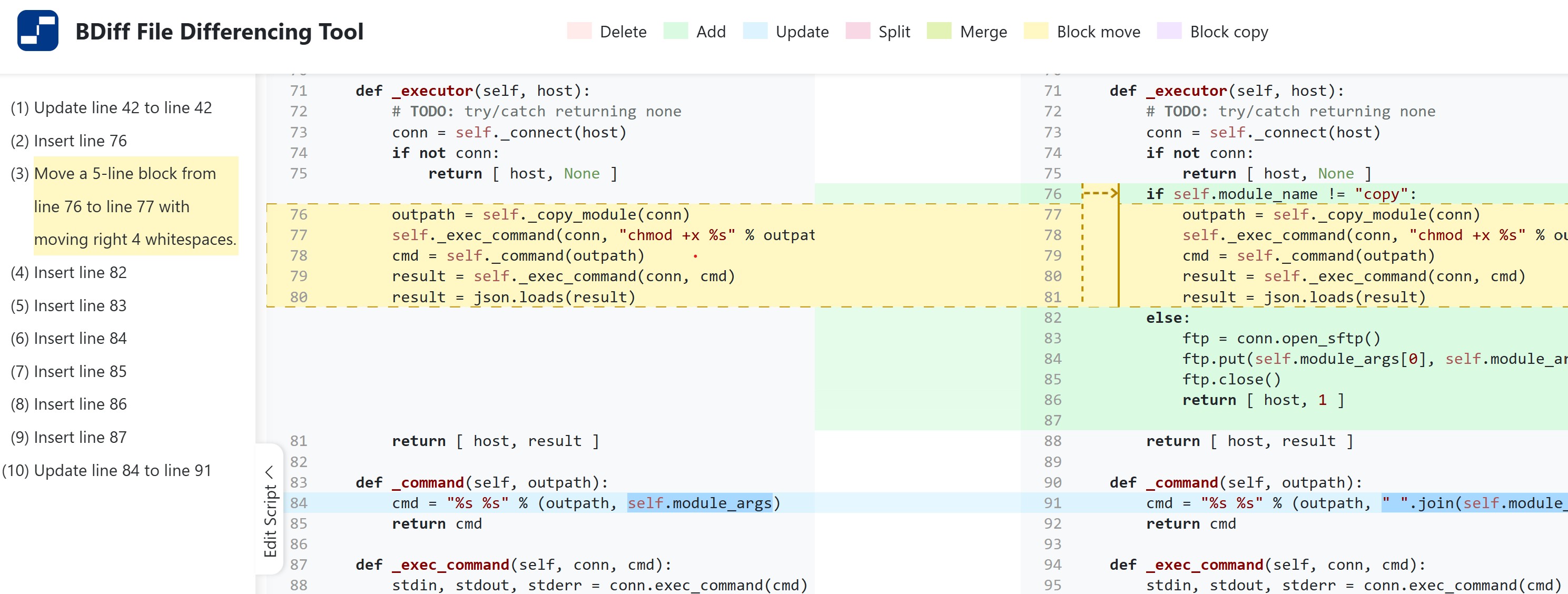}
    \caption{GitHub Project: ansible/ansible. Commit: 3807824. File: \_\_init\_\_.py}
    \label{addif}
    \vspace{8\baselineskip}
\end{figure}

\subsection{Duplicating and reusing an XML block}
\label{Reusing_XML_appendix}
\begin{figure}[H]
    \centering
    \includegraphics[width=\linewidth]{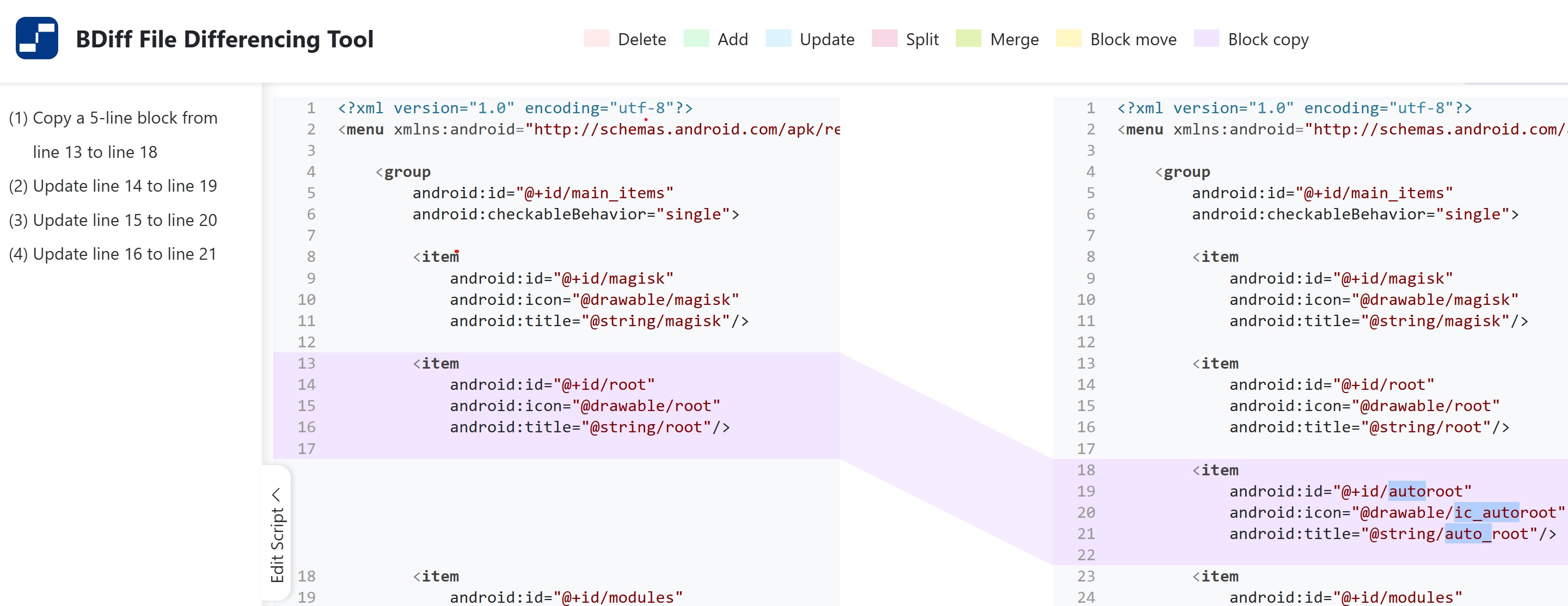}
    \caption{GitHub Project: topjohnwu/Magisk. Commit: fc5c964. File: drawer.xml}
    \label{dupxml}
\end{figure}

\subsection{Duplicating a Function Block into another Class}
\label{Reusing_func_appendix}
\begin{figure}[H]
    \centering
    \includegraphics[width=\linewidth]{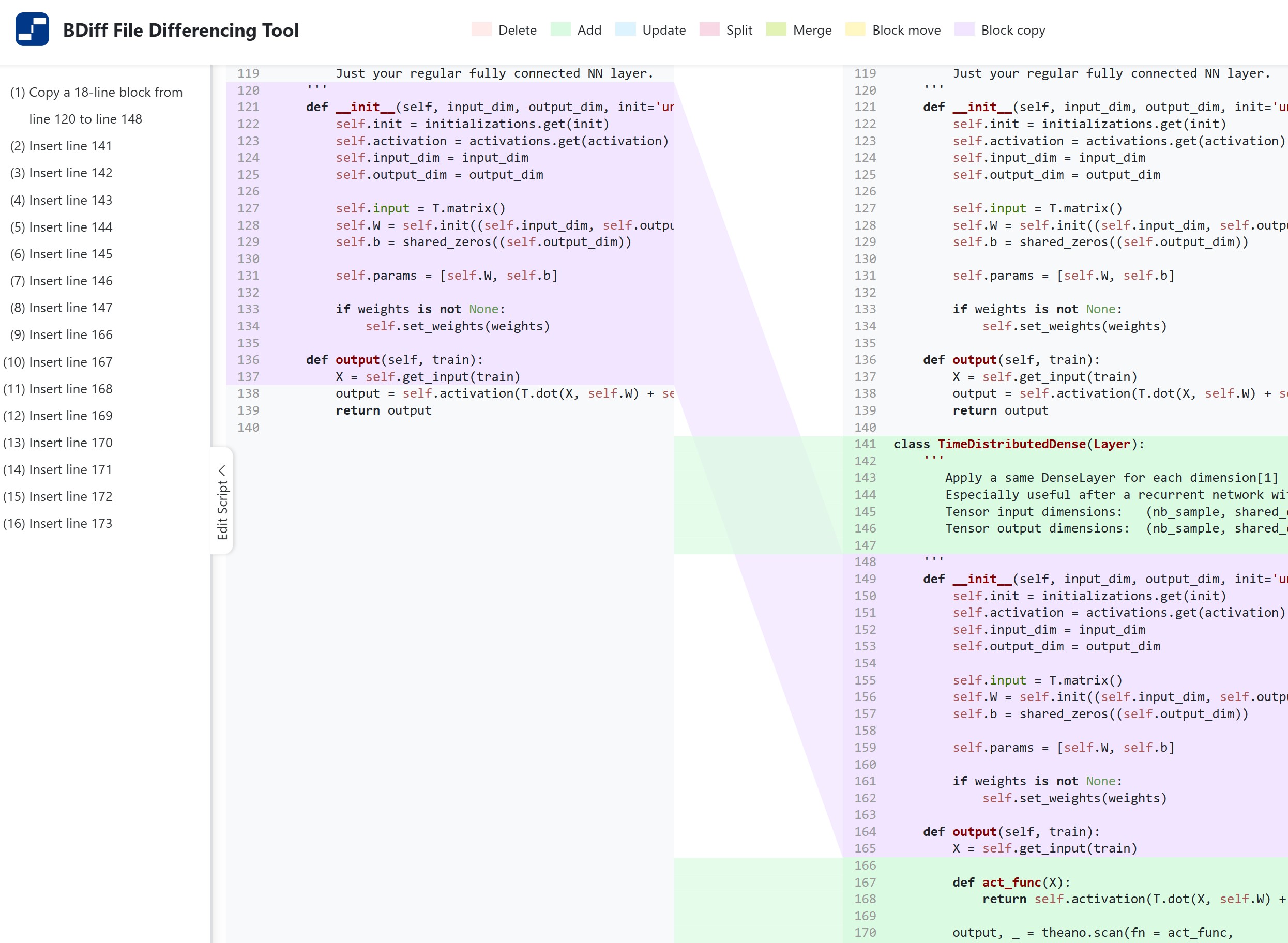}
    \caption{GitHub Project: keras-team/keras. Commit: aa7f9cd. File: core.py}
    \label{dupfuc}
\end{figure}

\subsection{Duplicating an XML Block into another Block with Indentation}
\label{Reusing_xml_appendix2}
\begin{figure}[H]
    \centering
    \includegraphics[width=\linewidth]{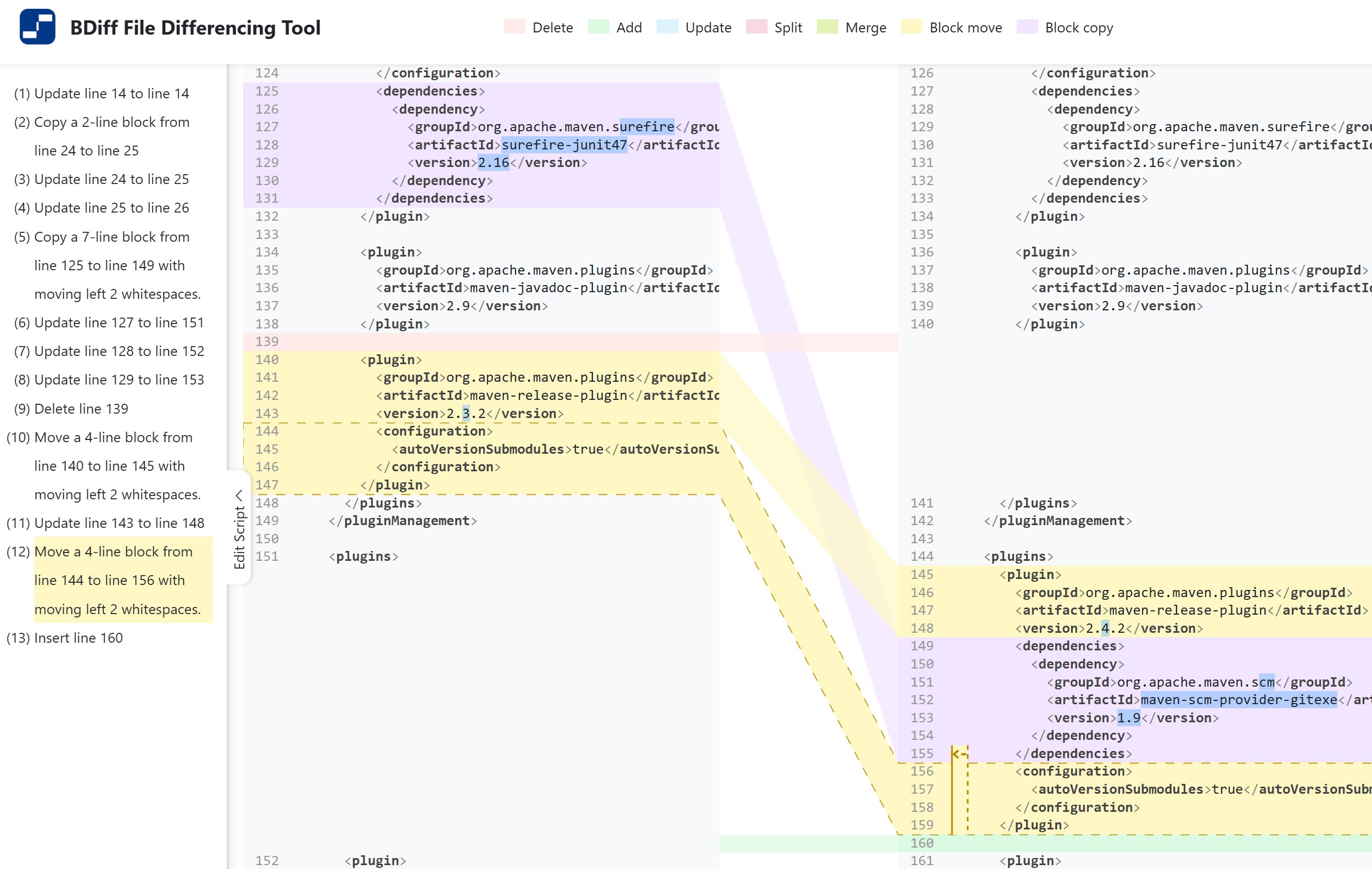}
    \caption{GitHub Project: square/okhttp. Commit: c863881. File: pom.xml}
    \label{copy_move}
\end{figure}

\subsection{Splitting Lines and Indenting Blocks}
\label{Splitting_appendix2}
\begin{figure}[H]
    \centering
    \includegraphics[width=\linewidth]{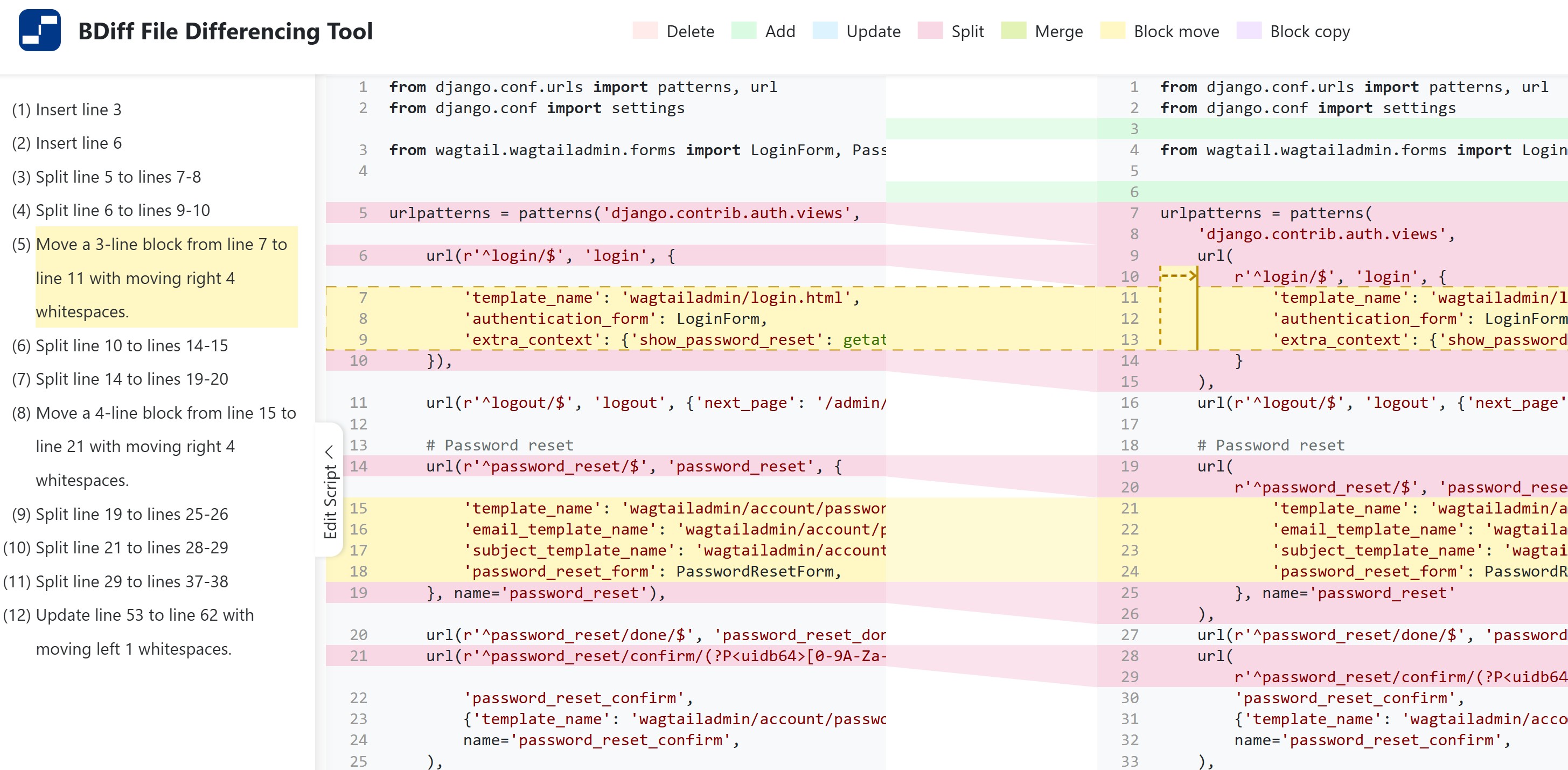}
    \caption{GitHub Project: wagtail/wagtail. Commit: a2a580f. File: urls.py}
    \label{split_idnent}
\end{figure}

\section{Typical Diff Cases of GumTree and \bdiff}
\subsection{GumTree Identifies an AST Node Change that Contains Multiple Lines as One EA}
\label{ast_change}
\begin{figure}[H]
\centering
\subfloat[\bdiff]{\includegraphics[width=\textwidth]{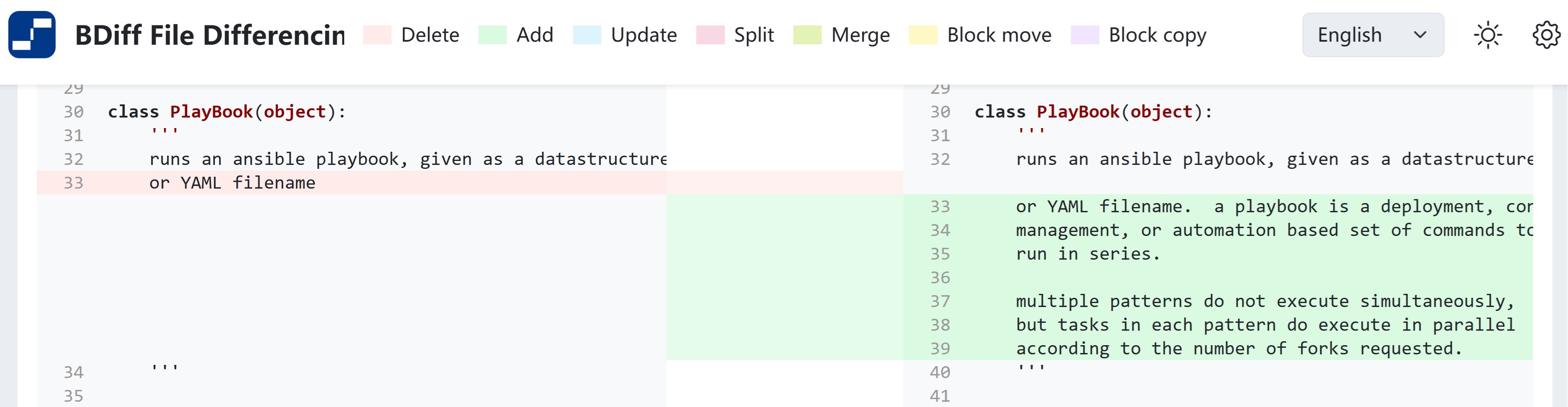}} \\
\subfloat[GumTree]{\includegraphics[width=\textwidth]{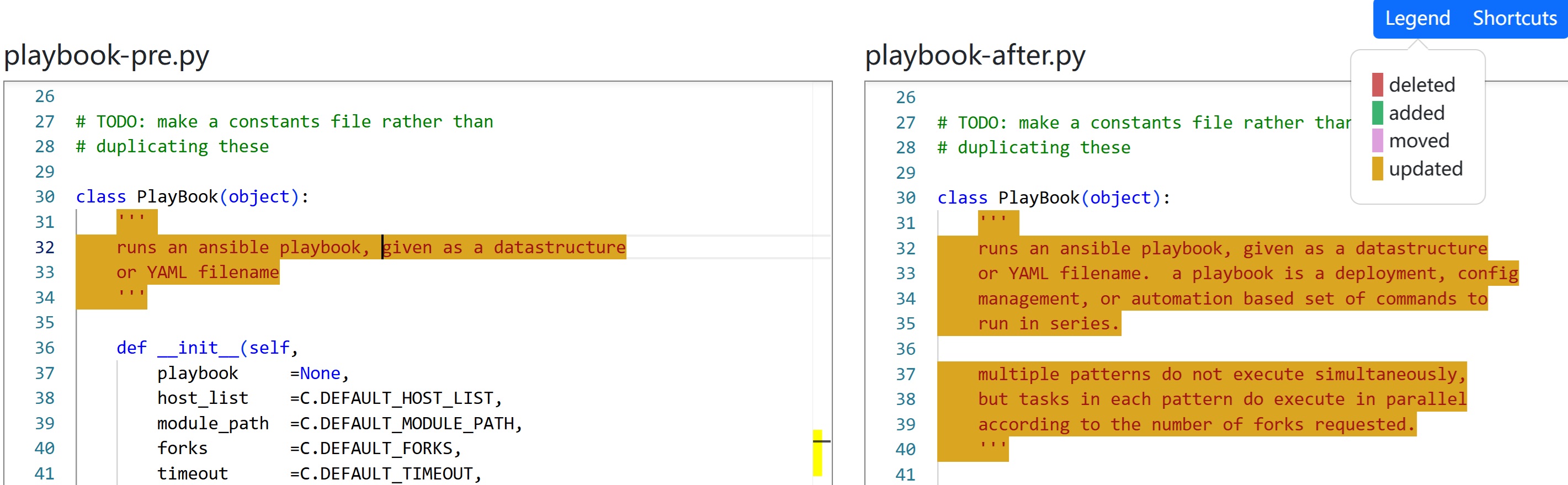}} \\
\caption{GitHub Project: ansible/ansible. Commit: 43f7dee. File: playbook.py}
\label{update_ast}
\vspace{19\baselineskip}
\end{figure}

\begin{figure}[H]
\centering
\subfloat[\bdiff]{\includegraphics[width=\textwidth]{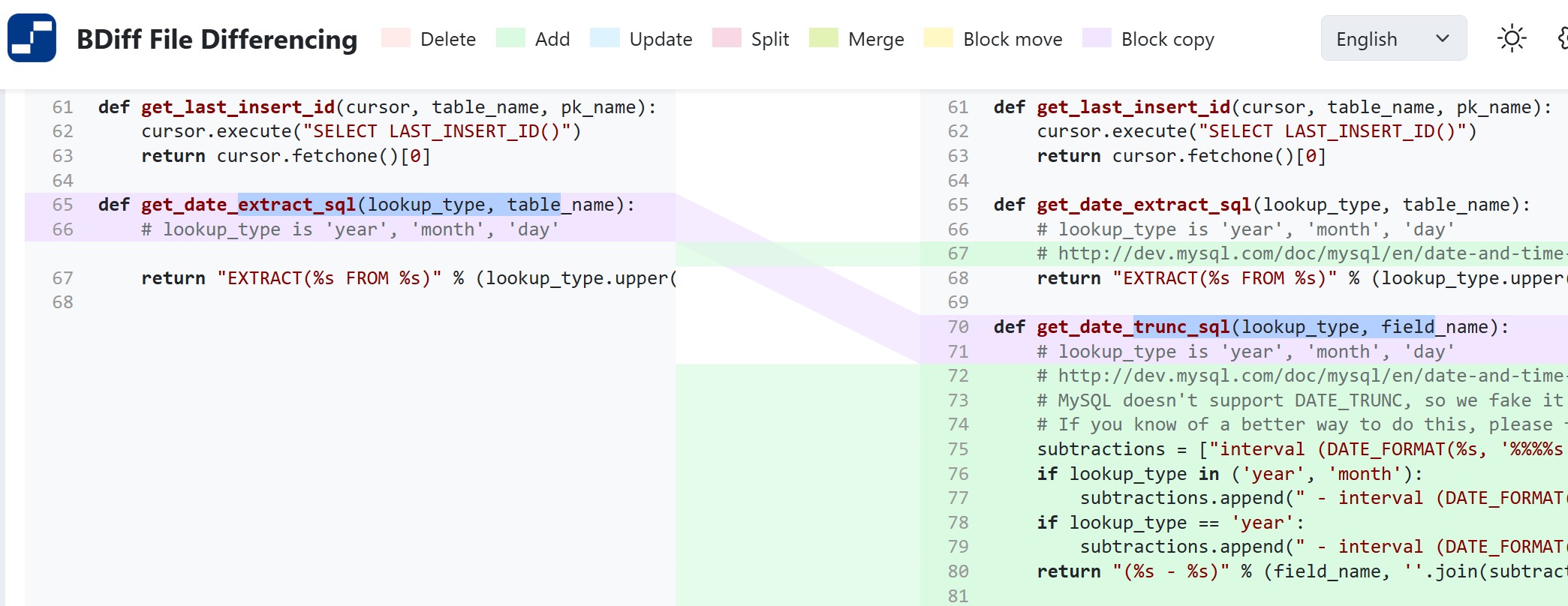}} \\
\subfloat[GumTree]{\includegraphics[width=\textwidth]{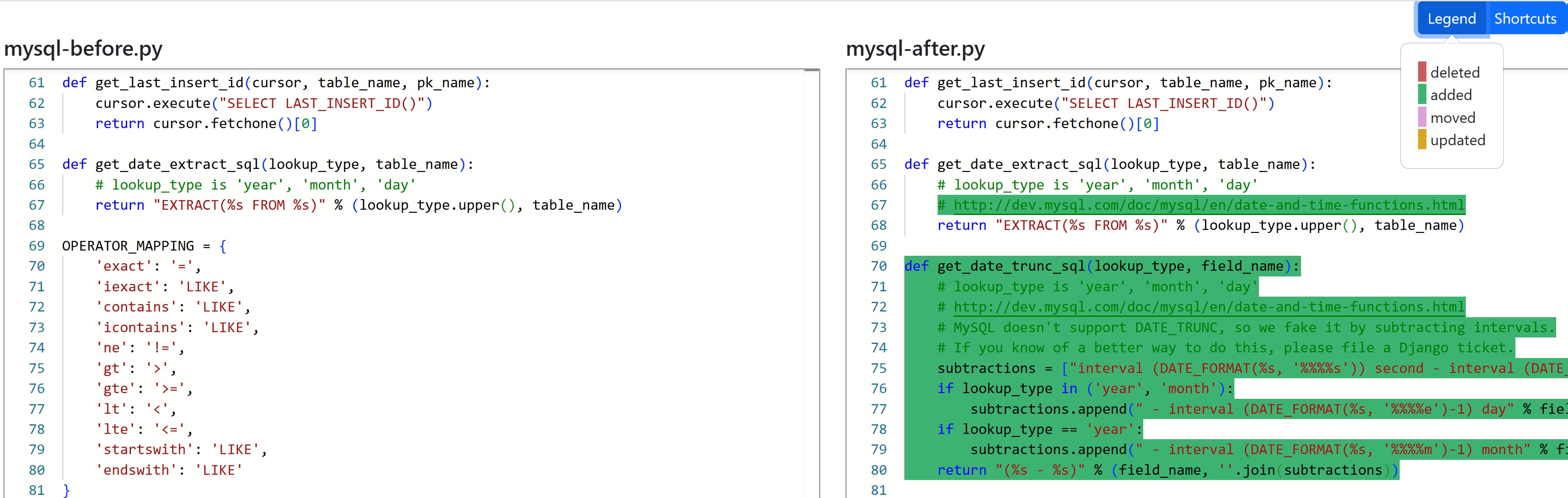}} \\
\caption{GitHub Project: django/django. Commit: b1c543d. File: mysql.py}
\label{insert_ast}
\end{figure}
\subsection{GumTree is Insensitive to Formatting Changes}
\label{gt_no_format}
\begin{figure}[H]
\centering
\subfloat[\bdiff]{\includegraphics[width=\textwidth]{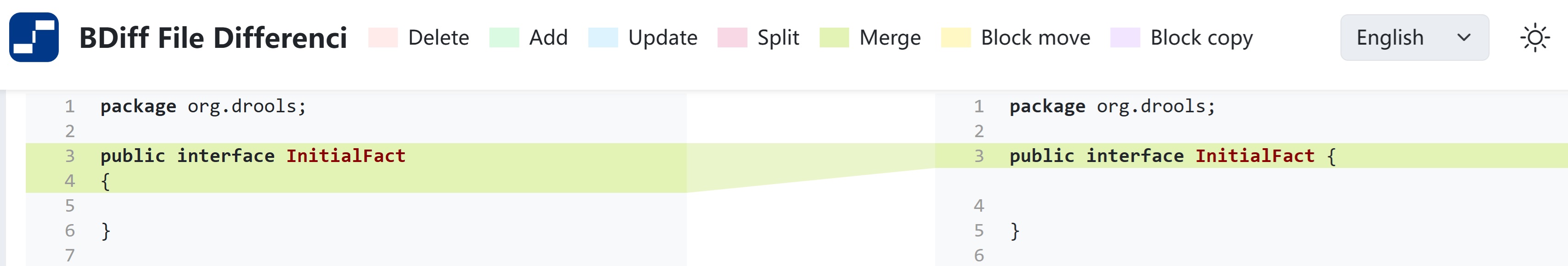}} \\
\subfloat[GumTree]{\includegraphics[width=\textwidth]{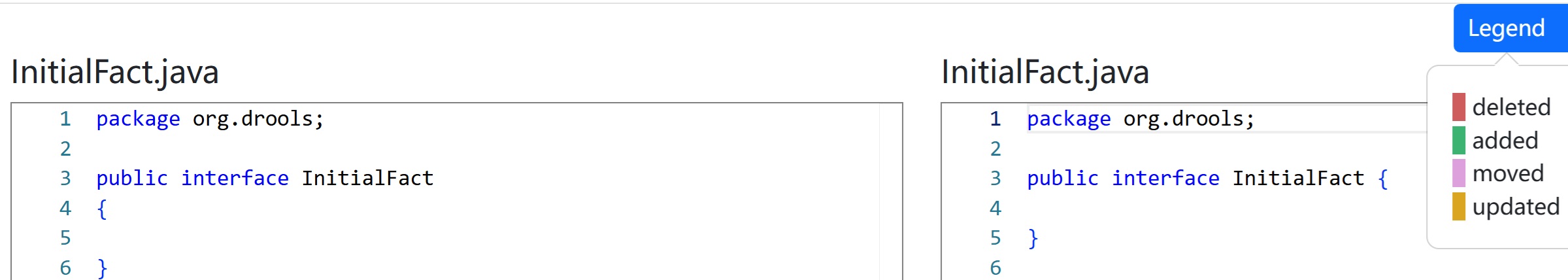}} \\
\caption{GitHub Project: apache/incubator-kie-drools. Commit: 5bb719f. File: InitialFact.java}
\label{no_format1}
\vspace{28\baselineskip}
\end{figure}

\begin{figure}[H]
\centering
\subfloat[\bdiff]{\includegraphics[width=\textwidth]{Figures/home-assitant_a2a580f0fe7a1354a109eb062b5393fbb330f508-urls.jpg}} \\
\subfloat[GumTree]{\includegraphics[width=\textwidth]{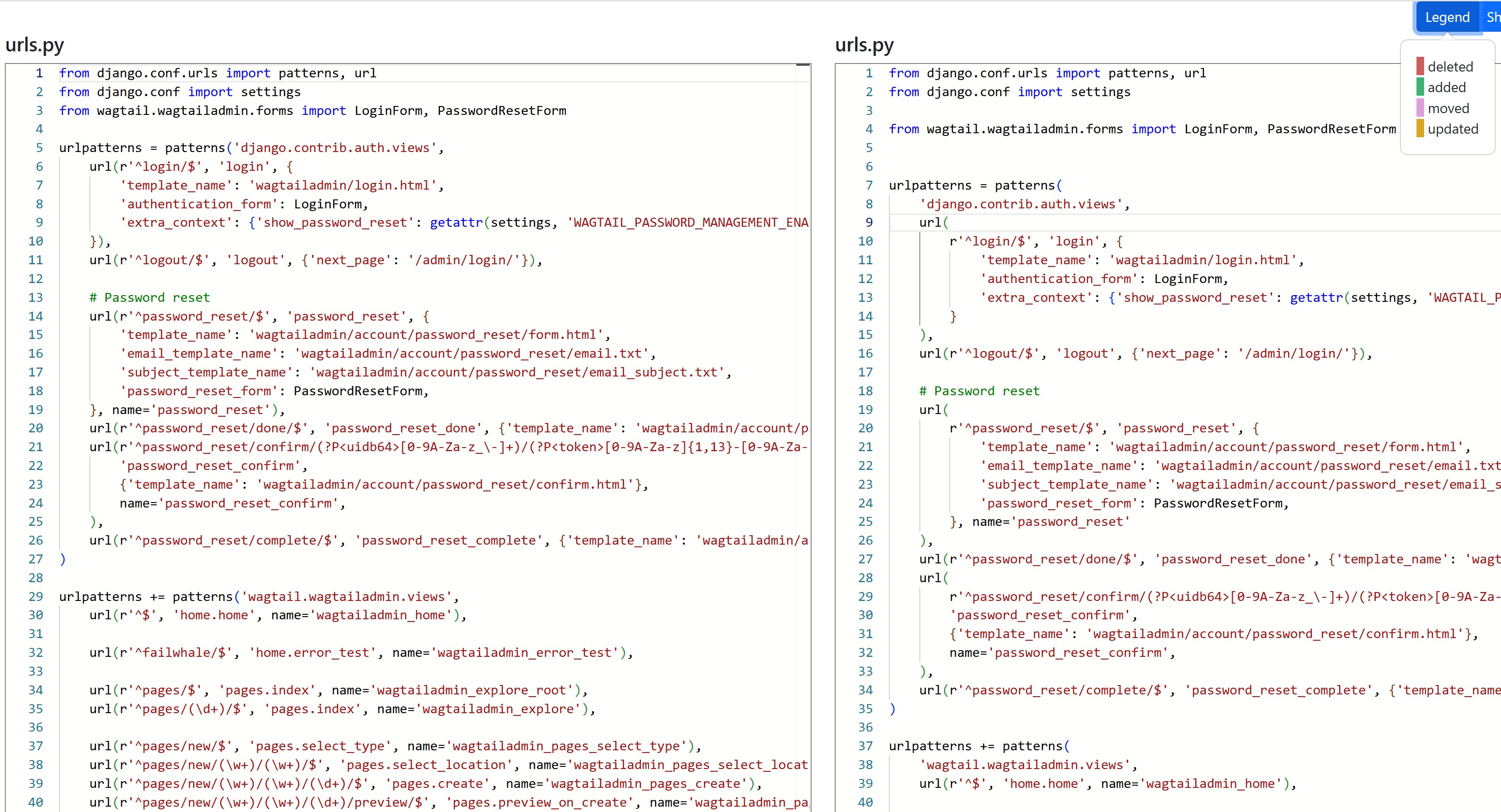}} \\
\caption{GitHub Project: wagtail/wagtail. Commit: a2a580f. File: urls.py}
\label{no_format2}
\end{figure}
\subsection{GumTree Identifies Consecutive Lines Moving as Individual Line Movements}
\label{gumtree_line_move_appendix}
\begin{figure}[H]
\centering
\subfloat[\bdiff]{\includegraphics[width=\textwidth]{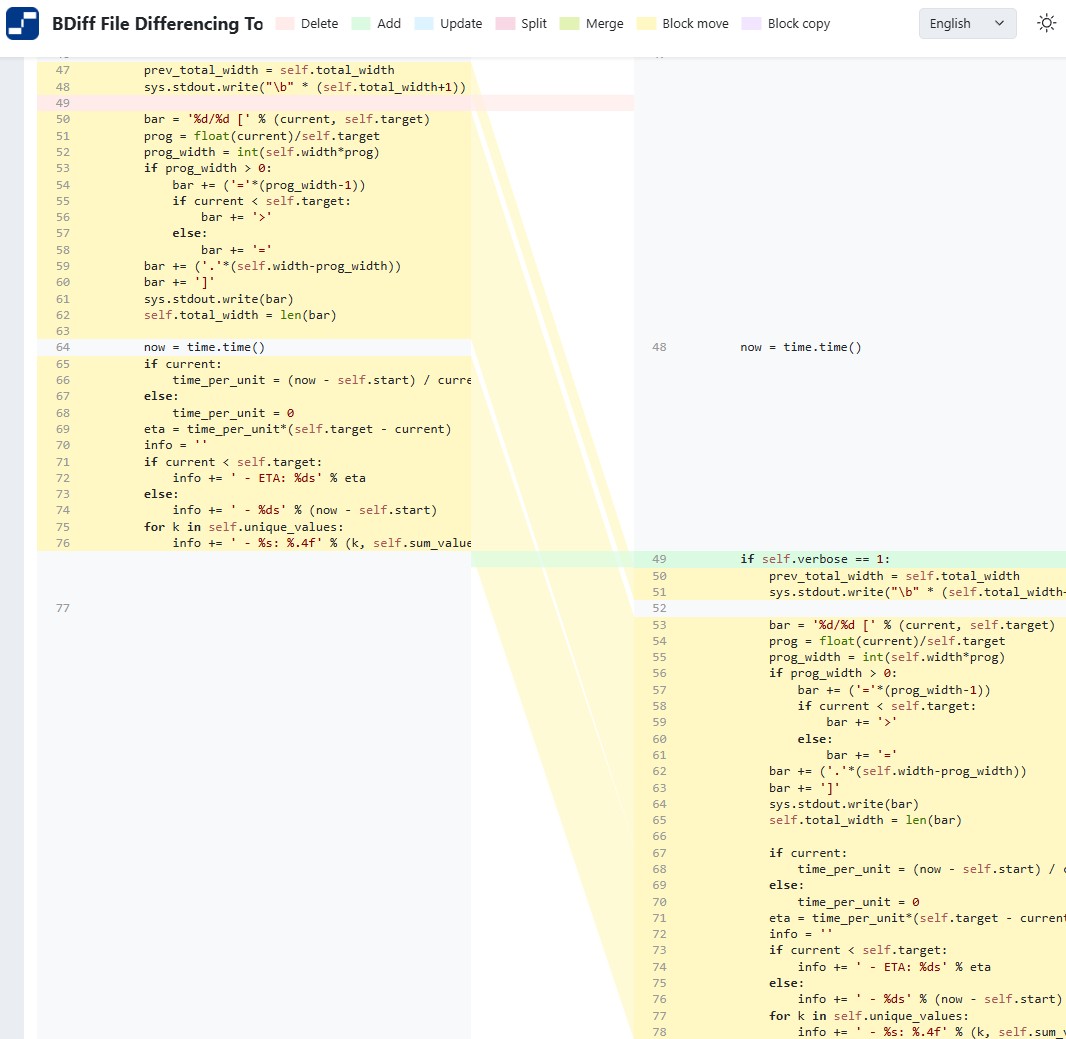}} \\
\label{individule_line_moving}
\vspace{11\baselineskip}
\end{figure}
\begin{figure}[H]
\ContinuedFloat
\centering
\subfloat[GumTree]{\includegraphics[width=\textwidth]{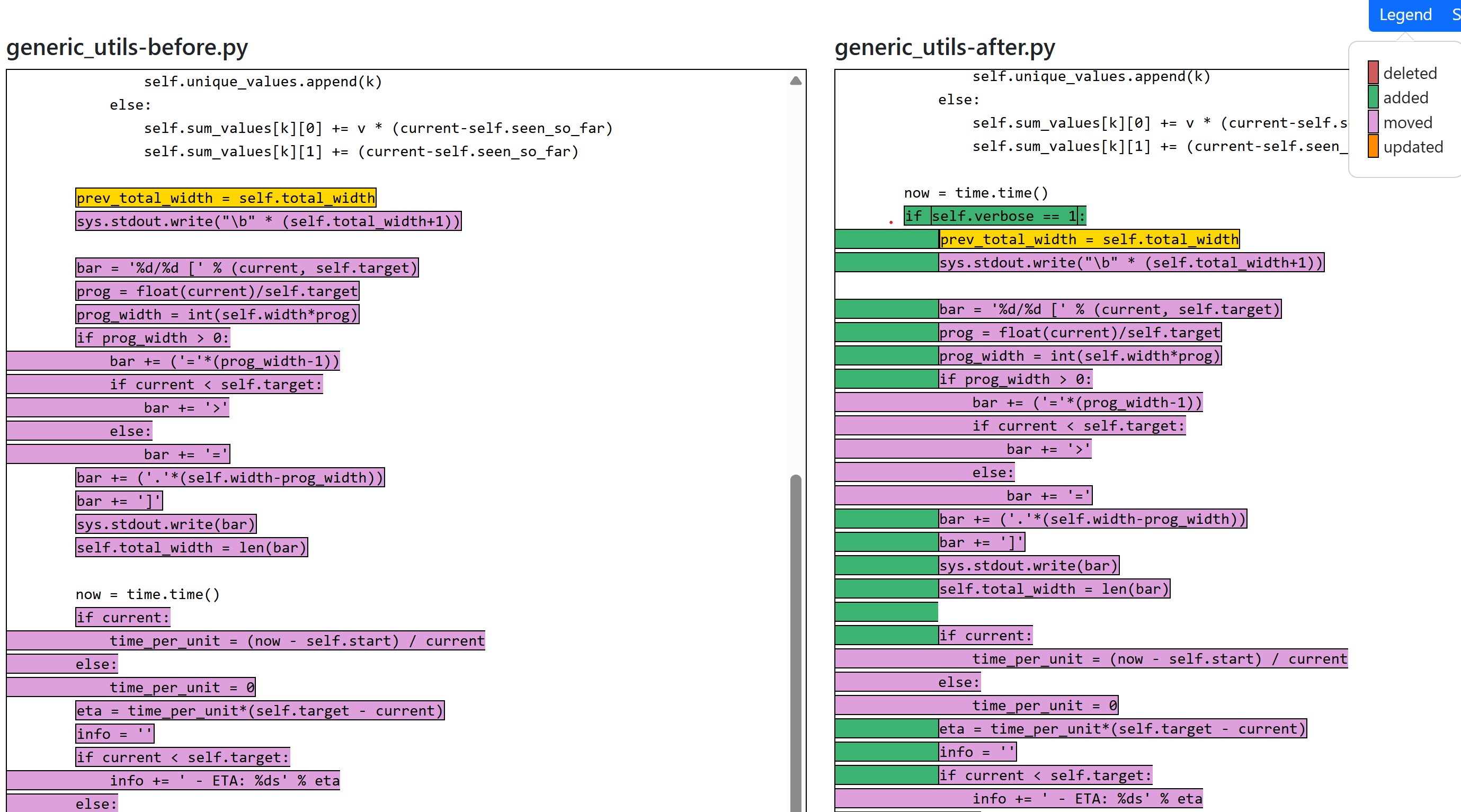}} \\
\caption{GitHub Project: keras-team/keras. Commit: c224482. File: generic\_utils.py}
\end{figure}

\end{document}